\documentclass[prd,,nofootinbib,12pt,showpacs,showkeys]{revtex4-1}
\usepackage{slashed}
\usepackage{amsmath}
\usepackage{amssymb}
\usepackage{graphicx} 
\usepackage{hyperref}
\usepackage{tikz}
\usetikzlibrary{arrows}
\usepackage{color}
\usepackage{bm}
\usepackage{bbold}
\usepackage{caption}
\usepackage{subcaption}

\usepackage{verbatim}
\usepackage{soul}

\newcommand{\be}{\begin{eqnarray}}
\newcommand{\ee}{\end{eqnarray}}

\newcommand{\bn}{\begin{enumerate}}
\newcommand{\en}{\end{enumerate}}

\newcommand{\bg}[1]{{\color{black} #1}}

\newcommand{\lsp}{\hspace{1pt}}

\def\Tr{{\rm Tr}}

\def\det{{\rm det}}

\def\vec#1{\bm{#1}}

\newcommand{\qed}{\nobreak \ifvmode \relax \else
      \ifdim\lastskip<1.5em \hskip-\lastskip
      \hskip1.5em plus0em minus0.5em \fi \nobreak
      \vrule height0.75em width0.5em depth0.25em\fi}

\allowdisplaybreaks[1]

\begin{document}
\title{On the Heat Kernel and Weyl Anomaly of  Schr\"odinger invariant theory}

\author{Sridip Pal}
\email{srpal@ucsd.edu}
\affiliation{Department of Physics, 
University of California, San Diego\\
La Jolla, CA 92093, USA}

\author{and Benjam\'\i{}n Grinstein}
\email{bgrinstein@ucsd.edu}
\affiliation{Department of Physics, 
University of California, San Diego\\
La Jolla, CA 92093, USA}

\begin{abstract}
  We propose a method inspired from discrete light cone quantization (DLCQ) to determine the heat
  kernel for a Schr\"odinger field theory (Galilean boost invariant with $z=2$ anisotropic scaling
  symmetry) living in $d+1$ dimensions, coupled to a curved Newton-Cartan background, starting from
  a heat kernel of a relativistic conformal field theory ($z=1$) living in $d+2$ dimensions. We use
  this method to show the Schr\"odinger field theory of a complex scalar field cannot have any Weyl
  anomalies. To be precise, we show that the Weyl anomaly $\mathcal{A}^{G}_{d+1}$ for Schr\"odinger
  theory is related to the Weyl anomaly of a free relativistic scalar CFT $\mathcal{A}^{R}_{d+2}$
  via $\mathcal{A}^{G}_{d+1}= 2\pi \delta (m) \mathcal{A}^{R}_{d+2}$ where $m$ is the charge of the
  scalar field under particle number symmetry. We provide further evidence of vanishing anomaly by
  evaluating Feynman diagrams in all orders of perturbation theory. We present an explicit
  calculation of the anomaly using a regulated Schr\"odinger operator, without using the null cone
  reduction technique. We generalise our method to show that a similar result holds for one time
  derivative theories with even $z>2$.
\end{abstract}

\pacs{04.62.+v,11.25.Hf}
\keywords{Non-relativistic CFT, Weyl Anomaly, Schr\"odinger Field theory, Heat Kernel}

\maketitle

\section{Introduction}
The Weyl anomaly in relativistic Conformal Field Theory (CFT) has a  rich
history \citep{Capper:1974ic,Deser:1976yx,Brown:1976wc,Dowker:1976zf,Hawking:1976ja,Christensen:1977jc,Duff:1977ay,Duff:1993wm}. In $1+1$ dimensions
irreversibility of RG flows has been established by Zamoldchikov
\citep{Zamolodchikov:1986gt} who showed  monotonicity of a quantity
$C$ that equals the Weyl anomaly
$c$ at fixed points. Remarkably, the anomaly $c$ equals the central
charge of the CFT. In
$3+1$ dimension, there is a corresponding ``$a$-theorem''
\citep{Osborn:1989td,Jack:1990eb, Komargodski:2011vj,
  Komargodski:2011xv} where $a$ again appears in the Weyl anomaly, and
there is strong evidence for a similar $a$-theorem in higher, even
dimensions \citep{Grinstein:2013cka,Grinstein:2014xba,Grinstein:2015ina,Stergiou:2016uqq}. In
contrast, much less is known in the case of non-relativistic field
theories admitting anisotropic scale invariance 
under the following transformation
\begin{align}
\vec{x}\to\lambda\vec{x},\qquad t\to \lambda^{z}t\,.
\end{align}
Nonetheless, non-relativistic conformal symmetry does emerge in
various scenarios. For example, fermions at unitarity, in which the
$S$-wave scattering length diverges, $|a|\rightarrow \infty$, exhibit
non-relativistic conformal symmetry. In ultracold atom gas
experiments, the $S$-wave scattering length can be tuned freely  along
an RG flow and this has renewed interest in the study of the RG flow
of such theories \citep{Regal:2004zza, Zwierlein:2004zz}. In fact, at
$a^{-1}=-\infty$ the system behaves as a BCS superfluid while at
$a^{-1}=\infty$ it becomes a BEC superfluid. The BCS-BEC crossover, at
$a^{-1}=0$, is precisely the unitarity limit, exhibiting
non-relativistic conformal symmetry
\citep{Nishida:2007pj,Nishida:2010tm}. In this regime, we expect
universality, with features independent of any microscopic details of
the atomic interactions. Other examples of non-relativistic systems
exhibiting scaling symmetry come with accidentally large scattering
cross section. Examples include various atomic systems,
like ${ }^{85}Rb$\citep{Roberts:1998zz},${}^{138}Cs$
\citep{Chin:2001uan}, and few nucleon systems like the deuteron
\citep{Kaplan:1998tg, Kaplan:1998we}.
 
Galilean CFT, which enjoys $z=2$ scaling symmetry is special among
Non-Relativistic Conformal Field Theories (NRCFTs). On group theoretic
grounds, there is a special conformal generator for $z=2$ that is not
present for $z\neq 2$
theories \citep{Balasubramanian:2008dm,Jensen:2014aia}. The coupling of
such theories to the Newton Cartan (NC) structure is well
understood \citep{Jensen:2014aia,
  Son:2013rqa,Geracie:2014nka,Perez-Nadal:2016tzr}. The generic discussion of
anomalies in such theories has been initiated by Jensen in
\citep{Jensen:2014hqa}. Moreover, there have been recent works
classifying and evaluating Weyl anomalies at fixed points \citep{Baggio:2011ha,
  Arav:2014goa, Arav:2016xjc, Arav:2016akx,Barvinsky:2017mal} and even away from the fixed points; the
latter have resulted in proposed $C$-theorem candidates \citep{Auzzi:2016lrq, Pal:2016rpz}.

It has been proposed in \citep{Jensen:2014hqa}, using the fact that
Discrete Light Cone Quantization (DLCQ) of a relativistic CFT living in
$d+2$ dimensions yields a non-relativistic Galilean CFT in $d+1$ dimensions with $z=2$, that
the Weyl anomaly of the relativistic CFT survives in the non-relativistic
theory. The conjecture states that the Weyl anomaly $\mathcal{A}^{G}$
for a Schr\"odinger field theory (Galilean boost invariant with
$z=2$ scale symmetry and special conformal symmetry) is given by
\begin{align}
\mathcal{A}^{G}_{d+1}=aE_{d+2}+\sum_{n}c_{n}W_{n}
\end{align}
where $E_{d+2}$ is the $d+2$ dimensional Euler density of the parent
space-time and  $W_{n}$ are Weyl covariant scalars with weight $(d+2)$. The right hand
side is computed on a geometry given in terms of the $d+2$ dimensional metric; this will
be explained below, see Eq.~\eqref{eq:backgroundmetric}. A specific example of particular
interest is 
\begin{align}
\mathcal{A}^{G}_{2+1}=aE_{4}-cW^{2}
\end{align}
where $W^{2}$ stands for  the square of the  Weyl tensor. 

The purpose of this work is twofold. First,  we show that these
proposed relations must be corrected to include a factor of
$\delta(m)$, when the Schr\"odinger invariant theory involves a single complex scalar
field having charge $m$ under the $U(1)$ symmetry. To be precise, we show that  
\begin{align}\label{mainresult}
\mathcal{A}^{G}_{d+1}=  2\pi \delta (m) \mathcal{A}^{R}_{d+2}
\end{align}
where $\mathcal{A}^{R}_{d+2}$ is the Weyl anomaly of the corresponding
relativistic CFT in $d+2$ dimensions. This is derived explicitly for the case of a bosonic (commuting) scalar
field, but the derivation applies equally to the case of a fermionic (anti-commuting) scalar field.
The second purpose is to develop a framework inspired from
DLCQ to evaluate the heat kernel of a theory with one time derivative
kinetic term in a non-trivial curved background. This framework
enables us to calculate not only the heat kernel but also the anomaly
coefficients. In fact, using this method and its appropriately
modified form enables us to generalise Eq.~\eqref{mainresult} to 
one time derivative theories with arbitrary even $z$, where the parent $d+2$
dimensional theory enjoys $SO(1,1) \times SO(d)$ symmetry with scaling
symmetry acting as $t\to\lambda^{z/2}t, x^{d+2}\to\lambda^{z/2}x^{d+2},
x^{i}\to\lambda x^{i}$, ($i=1,\ldots,d+1$). 

The paper is organised as follows. We will briefly review coupling of
a Schr\"odinger field theory to the Newton-Cartan structure in
Sec.~\ref{sec:NC}. In Sec.~\ref{DLCQ}, we sketch how DLCQ can be
used to obtain Schr\"odinger field theories following the procedure of
\citep{Jensen:2014hqa} and propose its modified cousin, that we call
Lightcone Reduction (LCR), to obtain a Schr\"odinger field
theory. In Sec.~\ref{heatkernel} we determine the heat kernel for free
Galilean CFT coupled to a flat NC structure in two different ways, on
the one hand using LCR and on the other without the use of DLCQ,
providing a check on our proposed method for determining the heat
kernel for Galilean field theory coupled to a curved NC geometry. We
then proceed to evaluate the heat kernel on curved spacetime according
to the proposal and subsequently derive the Weyl anomaly for
Schr\"odinger field theory of a single complex scalar. In Sec.~\ref{perturbation} we 
reconsider the computation using perturbation theory; we find that for a wide class of
models on a curved background all vacuum diagrams vanish.  In fact, we show that an
anomaly is not induced  in the more general case that $U(1)$ invariant
dimensionless couplings are included, regardless of
whether we are at a fixed point or away from it, in all orders of a
perturbative expansion in the dimensionless coupling and metric. In
Sec.~\ref{generalisation}, we give a formal proof of our prescription
and generalise the framework to calculate the heat kernel and anomaly
for theories with one time derivative and arbitrary even $z$. We
conclude with a brief summary of the results obtained and discuss future
directions of investigation. Technical aspects of defining heat kernel for one time
derivative theory in flat space-time are explored in App.~\ref{ap1}, and on a curved background in
App.~\ref{riemann}. Finally, in App.~\ref{hketa} we present an explicit
  calculation of the anomaly using a regulated Schr\"odinger operator, without using the null cone
  reduction technique. 

\section{Newton-Cartan Structure $\&$ Weyl Anomaly}\label{sec:NC}
The study of the Weyl anomaly necessitates coupling of non-relativistic
theory to a background geometry, which can potentially be curved.
Generically, the prescription for coupling to a background can depend on the global
symmetries of the theory on a flat background. Of interest to us are Galilean and
Schrodinger field theories. The algebra of the Galilean generators is given 
by~\citep{Balasubramanian:2008dm} 
\begin{gather}
[M_{ij},N]=0\,, \qquad [M_{ij}, P_{k}]=\imath
(\delta_{ik}P_{j}-\delta_{jk}P_{i})\,, \qquad [M_{ij},K_{k}]=\imath
(\delta_{ik}K_{j}-\delta_{jk}K_{i})\,,\nonumber \\
[M_{ij},M_{kl}]= \imath (\bg{\delta_{ik}M_{jl}}-\delta_{jk}M_{il}+\delta_{il}M_{kj}-\delta_{jl}M_{ki})\,,\nonumber \\
\label{eq:GalAlg1} 
[P_{i},P_{j}]=[K_{i},K_{j}]=0\,, \qquad [K_{i},P_{j}]=\imath\delta_{ij}N\,,\\
[H,N]=[H,P_{i}]=[H,M_{ij}]=0\,, \quad [H,K_{i}]=-\imath P_{i}\,, \nonumber  
\end{gather}
\bg{and the commutators of dilatation generator with that of Galilean ones are given by}
\begin{gather}
\qquad [D,P_{i}]=\imath P_{i}\,, \qquad [D,K_{i}]= (1-z)\imath K_{i}\,,  
\quad [D,H]=z\imath H\,,\nonumber \\
\label{eq:GalAlg2} 
\quad [D,N]=\imath (2-z)N\,, \quad [M_{ij},D]=0
\end{gather}
where $i,j =1,2, \ldots, d$ label the spatial dimensions, $z$ is the anisotropic exponent,
$P_{i}$, $H$ and $M_{ij}$ are generators of spatial translations, time translation spatial
rotations, respectively, $K_{i}$ generates Galilean boosts along the $x^{i}$ direction,
$N$ is the particle number (or rest mass) symmetry generator and $D$ is the generator of
dilatations. The generators of Schr\"odinger invariance include, in addition,  a generator
 of special conformal transformations, $C$. \bg{The Schr\"odinger algebra  consists of}  the $z=2$
 version of~\eqref{eq:GalAlg1},\eqref{eq:GalAlg2}  plus the commutators of $C$, 
\begin{align}
[M_{ij},C]=0\,, \qquad [K_{i},C]=0\,, \qquad [D,C]=-2\imath C\,, \qquad [H,C]=-\imath D.
\end{align}
In what follows, by  Schr\"odinger invariant theory we will mean a 
$z=2$ Galilean, conformally invariant theory. For $z\neq 2$ we only discuss anisotropic
scale invariant theories invariant under a group generated by  $P_{i}$,
$M_{ij}$, $H$, $D$ and $N$ such that the kinetic term involves one time derivative only. The most natural way to
couple Galilean (boost) invariant field theories to geometry is to use the
Newton-Cartan (NC) structure~\citep{Jensen:2014aia,
  Son:2013rqa,Geracie:2014nka}. In what follows we briefly review NC
geometry, following Ref.~\citep{Jensen:2014hqa}. 

The NC structure defined on a $d+1$ dimensional manifold
$\mathcal{M}_{d+1}$ consists of a one form $n_{\mu}$, a symmetric
positive semi-definite rank $d$ tensor $h_{\mu\nu}$ and an $U(1)$
connection $A_{\mu}$, such that  the metric tensor
\begin{align}
\label{eq:Themetric}
g_{\mu\nu}=n_{\mu}n_{\nu}+h_{\mu\nu} 
\end{align}
is positive definite. The upper index data $v^{\mu}$ and $h^{\mu\nu}$ is defined by
\begin{align}
\label{eq:Milneconds}
v^{\mu}n_{\mu}=1,\qquad v^{\nu}h_{\mu\nu}=0,\qquad h^{\mu\nu}n_{\nu}=0,\qquad h^{\mu\rho}h_{\rho\nu}=\delta^{\mu}_{\nu}-v^{\mu}n_{\nu}
\end{align}
Physically $v^{\mu}$ defines a local time direction while $h_{\mu\nu}$ defines a metric on spatial slice of $\mathcal{M}_{d}$. 

As prescribed in \citep{Jensen:2014aia}, while coupling a Galilean
invariant field theory to a  NC structure, we demand 
\begin{enumerate}\label{list}
\item Symmetry under reparametrization of co-ordinates. Technically,
  this requirement boils down to writing the theory in a diffeomorphism
  invariant way. 
\item $U(1)$ gauge invariance. The fields belonging to some
  representation of Galilean algebra carry some charge under particle
  number symmetry, which is an $U(1)$ group. Promoting this to a local
  symmetry   requires a gauge field $A_{\mu}$ that
  is  sourced by the $U(1)$ current.  
\item Invariance under Milne boost  under which  $(n_{\mu},h^{\mu\nu})$ remains invariant, while 
\begin{align}\label{boostsymmetry}
v^{\mu}\to v^{\mu}+\psi^{\mu}, \quad h_{\mu\nu}\to h_{\mu\nu}- \left(n_{\mu}\psi_{\nu}+n_{\nu}\psi_{\mu}\right)+n_{\mu}n_{\nu}\psi^{2}, \quad A_{\mu}\to A_{\mu}+\psi_{\mu}-\frac{1}{2}n_{\mu}\psi^{2}
\end{align}
where $\psi^{2}=h^{\mu\nu}\psi_{\mu}\psi_{\nu}$ and $v^{\nu}\psi_{\nu}=0$.
\end{enumerate}

The action of a free Galilean scalar $\phi_m$  with charge $m$,
coupled to this NC structure satisfying all the symmetry conditions
listed above is given by  
\begin{align}
\int d^{d+1}x \sqrt{g}\left[\imath m v^{\mu} \left(\phi_{m}^{\dagger}D_{\mu}\phi_m-\phi_{m}D_{\mu}\phi_{m}^{\dagger}\right)-h^{\mu\nu}D_{\mu}\phi_{m}^{\dagger}D_{\nu}\phi_{m}\right]
\end{align}
where $D_{\mu}=\partial_{\mu}-\imath m A_{\mu}$ is the appropriate gauge invariant derivative.

From a group theory perspective, a Galilean group can be a  subgroup
of a larger group that includes dilatations. That is, besides the
symmetries mentioned earlier, a Galilean invariant field theory
coupled to the flat NC structure can also be scale invariant, {\it
  i.e.,} invariant under the following transformations
\begin{align}
\vec{x} \to \lambda \vec{x}, \qquad t \to \lambda^{z}t,
\end{align}
where $z$ is the  dynamical critical exponent of the theory. As mentioned earlier, for
$z=2$, the symmetry algebra may further be enlarged to contain a special
conformal generator,  resulting in the  Schr\"odinger
group. On coupling a
Galilean CFT with arbitrary $z$ to a nontrivial curved NC structure, the scale invariance
can be thought of as invariance under following scaling of NC data
(also known as anisotropic Weyl scaling; henceforth we omit the word
\textit{anisotropic}, and by Weyl transformation it should be
understood that we mean the
transformation with appropriate~$z$): 
\begin{align}\label{weylvariation}
n_{\mu}\to e^{z\sigma}n_{\mu},\qquad h_{\mu\nu}\to e^{2\sigma}h_{\mu\nu}, \qquad A_{\mu}\to e^{(2-z)\sigma}A_{\mu},
\end{align}
where $\sigma$ is a function of space and time. 

Even though classically a Galilean CFT may be scale invariant, it is
not necessarily true that it remains invariant quantum mechanically.
Renormalisation may lead to anomalous breaking of scale symmetry much
like in the  Weyl anomaly in relativistic CFTs (where $z=1$).
The anomaly $\mathcal{A}$ is defined from the infinitesimal Weyl variation
\eqref{weylvariation} of the connected generating functional $W$:
\begin{align} 
\delta_{\sigma}W= \int d^{d+1}x\,\sqrt{g}\,  \delta\sigma\, \mathcal{A},.
\end{align}

We mention in passing that away from the fixed point the coupling is
scale dependent, that is, the running of the coupling under the RG
must be accounted for, hence the variation $\delta_{\sigma}$ on the
couplings needs to be incorporated. The generic scenario has been
elucidated in Ref.~\citep{Pal:2016rpz}. 

In this work, we are interested
in anomalies at a fixed point. Even in the absence of running of the
coupling, the  background
metric can act as an external operator insertion on vacuum bubble
diagrams leading to new UV divergences that are absent  in
flat space-time. Removing these new divergences can potentially lead
to anomalies. The anomalous ward identity for anisotropic Weyl
transformation is given by\citep{Jensen:2014hqa}
\begin{equation}
\label{anomaly}
zn_{\mu}\mathcal{E}^{\mu}-h^{\mu\nu}T_{\mu\nu}= \mathcal{A}\,,
\end{equation}
where $n_{\mu}\mathcal{E}^{\mu}$ and $h^{\mu\nu}T_{\mu\nu}$ are
respectively diffeomorphic invariant measure of energy density and
trace of spatial stress-energy tensor.

In what follows, we will be interested in evaluating the quantity
appearing on the right hand side of Eq.~\eqref{anomaly}. A standard method is
through the evaluation of the heat kernel in a curved background.
Hence, our first task is to figure out a way to obtain the heat kernel for theories with
kinetic term involving only one time derivative. In the next  few
sections  we will introduce methods for computing heat kernels and arrive at the same
result from different approaches.

\section{Discrete Light Cone Quantization (DLCQ) $\&$ its cousin
  Lightcone Reduction (LCR)}
\label{DLCQ} 
One elegant way to obtain the heat kernel is to use Discrete
Light Cone Quantization (DLCQ). This  exploits the well known fact that a
$d+1$ Galilean invariant field theory can be constructed by starting
from a relativistic theory in $d+2$ dimensional Minkowski space in
light cone coordinates 
\begin{align}
ds^{2}=2dx^{+}dx^{-}+dx^{i}dx^{i}
\end{align}
where $i=2,3,\ldots, d+1$ and $x^{\pm}=\frac{x^{1}\pm t}{\sqrt{2}}$
define light cone co-ordinates, followed by a compactification in
the null co-ordinate $x^{-}$ on a circle. From here on, by
\textit{reduced} theory we will mean the theory in $d+1$ dimensions
while by \textit{parent} theory we will mean the $d+2$ dimensional
theory on which this DLCQ trick is applied. We  first present 
a brief review of DLCQ. 

The generators of $SO(d+1,1)$ which commute with $P_{-}$, the
generator of translation in the $x^{-}$ direction, generate the
Galilean algebra. $P_{-}$ is interpreted as the generator of particle
number of the reduced theory. In light cone
coordinates the mass-shell condition for a massive particle becomes\footnote{\bg{The unusual
  sign convention in our definition of $x^-$ results in the peculiar sign in Eq.~\eqref{massshellcondition}.}}
\begin{align}\label{massshellcondition}
p_{+}=\frac{|\vec{p}|^{2}}{2(-p_{-})}+\frac{M^{2}}{4(-p_{-})}
\end{align}
\bg{Eq.~\eqref{massshellcondition} can be interpreted  as the non-relativistic energy  of a particle, $p_+$,  with mass $m= -p_{-}$ in a  constant potential. The reduced mass-shell condition \eqref{massshellcondition} is Galilean invariant, that is, invariant  under boosts ($\vec{v}$) and rotations ($\mathbf{R}$):
\begin{align}
\nonumber \vec{p} \to \mathbf{R}\vec{p}-\vec{v} p_-, \qquad p_{+} \to p_+ +\vec{v}\cdot \left(\mathbf{R}\vec{p}\right)-\frac{1}{2}|\vec{v}|^2p_-
\end{align}
}
Setting $M=0$, the dispersion relation is of the form 
\begin{equation}
\omega=\frac{k^{2}}{2m}
\end{equation}
and enjoys $z=2$ scaling symmetry. \bg{To rephrase, setting $M=0$ will allow one to append a dilatation generator, which acts as follows:
\begin{align}
\nonumber p_{+} \to \lambda^2p_{+}\,,\quad p_{-} \to p_-, \quad \vec{p} \to \lambda \vec{p}
\end{align}}
Had we not compactified in the $x^-$ direction, $p_{-}$ would be
a continuous variable. The parameter $p_{-}$ can be changed using a boost in the ${+-}$ direction,
but compactification in the $x^{-}$ direction spoils relativistic boost symmetry and the eigenvalues
of $p_{-}$ become discretized, $p_{-}=\frac{n}{R}$, where $R$ is the compactification radius. We
note that Lorentz invariance is recovered in the $R\to\infty$ limit.  For convenience, by appropriately
rescaling the generators of spatial translations and of special conformal transformations, as well
as $P_-$, we can set $R=1$.

One can technically perform DLCQ even in a curved space-time as long as the metric admits
  a null isometry. This guarantees that we can adopt a coordinate system with a null coordinate
  $x^{-}$ such that all the metric components are independent of $x^{-}$. To be specific, we will
  consider the following metric:
\begin{align}
\label{eq:backgroundmetric}
ds^{2}=G_{MN}dx^{M}dx^{N}, \qquad G_{\mu -}= n_{\mu}, \qquad G_{\mu\nu}=h_{\mu\nu}+n_{\mu}A_{\nu}+n_{\nu}A_{\mu}, \qquad G_{--}=0
\end{align}
where $M,N=+,-, 1,2,\ldots,d$ run over all the indices in $d+2$ dimensions, the index
$\mu=+,1,2,\ldots,d$ runs over $d+1$ dimensions and $h_{\mu\nu}$ is a rank $d$ tensor. Ultimately,
$h_{\mu\nu}, n_{\mu}, A_{\mu}$ are to be identified with the NC structure, and just as above  we can
construct $h^{\mu\nu}$ and $v^{\mu}$ such that Eq.~\eqref{eq:Milneconds} holds. Moreover, these
quantities transform under Milne  boost symmetry as per Eq.~\eqref{boostsymmetry}. Hence, the boost invariant inverse metric is given by 
\begin{equation}\label{inversemetric}
G^{-\mu}=v^{\mu}-h^{\mu\nu}A_{\nu}, \qquad G^{\mu\nu}=h^{\mu\nu}, \qquad G^{--}= -2 v^{\mu}A_{\mu} + h^{\mu\nu}A_{\mu}A_{\nu}\,.
\end{equation}
Reduction on  $x^{-}$ yields a Galilean
invariant theory coupled to an NC structure given by
$(n_{\mu},h^{\mu\nu},A_{\mu})$, with metric given by~\eqref{eq:Themetric}. Moreover, all the symmetry
requirements listed above Eq.~\eqref{boostsymmetry} are satisfied by construction. 

This prescription allows us to construct Galilean QFT coupled to a non trivial NC structure
starting from a relativistic QFT placed in a curved background with one
extra dimension.  For example, we can consider DLCQ of a conformally
coupled scalar field in $d+2$ dimensions,  
\begin{align}\label{actionr}
S_{R}=\int d^{d+2}x \sqrt{-G} \left[-G^{MN}\partial_{M}\Phi^{\dagger}\partial_{N}\Phi - \xi \mathcal{R}\Phi^{\dagger}\Phi\right]\,, \qquad\xi=\frac{d}{4(d-1)}
\end{align}
where $\mathcal{R}$ stands for the Ricci scalar corresponding to the $G_{MN}$ metric.  
We compactify $x^{-}$ with periodicity $2\pi$ and expand $\Phi$ in fourier modes as
\begin{align}
\Phi=\frac{1}{\sqrt{2\pi}}\sum_{m} \phi_{m}(x^{\mu})e^{\imath m x^{-}}, \qquad \phi_{m} = \frac{1}{\sqrt{2\pi}}\int_{0}^{2\pi}dx^{-}\ \Phi e^{-\imath m x^{-}}\,.
\end{align}
In terms of $\phi_{m}$ , we recast the action, Eq.~\eqref{actionr} in following form using Eq.~\eqref{inversemetric}
\begin{align}\label{actiong}
S_{R}= \sum_{m} \int d^{d+1}x \sqrt{g}\left[\imath m v^{\mu}
\left(\phi_{m}^{\dagger}D_{\mu}\phi_m-\phi_{m}D_{\mu}\phi_{m}^{\dagger}\right)
-h^{\mu\nu}D_{\mu}\phi_{m}^{\dagger}D_{\nu}\phi_{m}-\xi\mathcal{R}\phi_{m}^{\dagger}\phi_{m}\right]
\end{align}
where $D_{\mu}=\partial_{\mu}-\imath m A_{\mu}$ and  where each of the $\phi_{m}$ carry  charge $m$ under the particle
number symmetry and sit in distinct  representations of the Schr\"odinger group.
The theory described by Eq.~\eqref{actiong} is
not Lorentz invariant because we have a discrete sum over $m$,
breaking the boost invariance along the null direction. 

The point of DLCQ is to break Lorentz invariance to Galilean
invariance. As explained above, one can work in the uncompactified limit,
and still break the Lorentz invariance by dimensional reduction. In
the uncompactified limit, the sum over eigenvalues of $P_-$ becomes integration over the continuous variable $p_{-}$. Nonetheless,
one can focus on any particular Fourier mode.  Technically, we can
implement this by performing a Fourier
transformation with respect to $x^{-}$ of quantities of interest. This procedure  also yields
a Galilean invariant field theory where the elementary field is the
particular Fourier mode under consideration. Henceforth we will refer
to this modified version of DLCQ as Lightcone Reduction (LCR). 

Taking a cue from the relation between the actions given by
Eqs.~\eqref{actionr} and \eqref{actiong} we propose the following
prescription to extract the heat kernel in the reduced theory:  

\noindent \textit{The heat kernel operator $K_{G}$ in $d+1$ dimensional Galilean
  theory is related to the heat kernel operator $K_{R}$ of the  parent
  $d+2$ dimensional relativistic theory via} 
\begin{equation}
\label{eq:prescription}
\langle (\vec{x}_2,t_2)|K_{G}|(\vec{x}_1,t_1)\rangle = 
\int_{-\infty}^{\infty} dx^{-}\ \langle \vec{x}_{2}, x^{-}_{2}, x_{2}^{+}|K_{R}|\vec{x}_{1}, x^{-}_{1}, x_{1}^{+}\rangle \ e^{-\imath m x^{-}_{12}}
\end{equation}
\textit{where $x^{-}_{12}=x^{-}_{2}-x^{-}_{1}$ and the time $t$ in
the reduced theory is to be equated with $x^{+}$ in the parent theory.} 

We will postpone the proof of our prescription to
Sec.~\ref{generalisation}. In the next section, we will lend
support to our prescription by verifying our
claim using two different methods of calculating the  heat kernel. \bg{We emphasize that the reduction prescription, described above, is applicable to  the $z=2$ case of  Galilean and scale invariant theories. The generic reduction procedure for arbitrary $z$ (though not Galilean boost invariant) is discussed later in sec.~\ref{sec:gener}.}
 
\section{Heat Kernel for a Galilean CFT with $z=2$}
\label{heatkernel}

\subsection{Preliminaries: Heat Kernel, Zeta Regularisation}
We start by briefly reviewing the heat kernel and zeta function regularisation method
\citep{Jack:1983sk, jack1986renormalizability, Jack:1990eb, Grinstein:2015ina}. A
pedagogical discussion can be found in \citep{mukhanov2007introduction,
  Vassilevich:2003xt}. Let us consider a theory with partition function $\mathcal{Z}$,
formally given by
\begin{align}
\mathcal{Z}= \int\, [\mathcal{D}\phi]  [\mathcal{D}\phi^\dagger]\, e^{-\int d^{d}x\, \phi^{\dagger}\mathcal{M}\phi}
\end{align}
where the eigenvalues of the operator $\mathcal{M}$  have positive
real part.\footnote{Positivity is required for convergence of the  gaussian integral.}The path integral over
the field variable $\phi$ suffers from ultraviolet (UV) divergences
and requires proper regularization and renormalisation to be rendered
as a meaningful finite quantity. Similarly, the quantum effective
action $W=-\ln\mathcal Z$ corresponding to this theory, given by a formal expression
\begin{align*}
W = \ln (\det(\mathcal{M}))
\end{align*}
requires regularization and renormalisation.\footnote{For anti-commuting fields $W =- \ln
    (\det(\mathcal{M}))$; the minus sign is the only difference between commuting and anti-commuting
  cases,  so that in what follows we restrict our attention to the case of commuting fields.}

The method of  zeta-function regularization introduces several  
quantities; the heat kernel operator
\begin{align}\label{heatkerneloperator}
\mathcal{G}= e^{-s\mathcal{M}}\,,
\end{align}
 its trace $K$ over the space $L^2$ of square integrable functions
\begin{align}\label{eq:Kdefined}
K(s,f, \mathcal{M})= \Tr_{L^{2}}\left(f\mathcal{G}\right)= \Tr_{L^{2}} \left(fe^{-s\mathcal{M}}\right)\,,
\end{align}
where $f\in L^2$, and the  zeta-function, defined as
\begin{align}\label{eq:zetadefined}
\zeta (\epsilon , f, \mathcal{M} ) = \Tr_{L^{2}} \left(f \mathcal{M}^{-\epsilon}\right)\,.
\end{align}
$K$ and $\zeta$ are related via  Mellin transform, 
\begin{equation}\label{eq:mellintransf}
K(s,f, \mathcal{M})= \frac{1}{2\pi\imath}\int_{c-\imath\infty}^{c+\imath\infty} \!\!\!\!d\epsilon\
s^{-\epsilon} \Gamma(\epsilon) \zeta(\epsilon, f, \mathcal{M})\quad
\text{and}\quad
\zeta(\epsilon, f, \mathcal{M})=
\frac1{\Gamma(\epsilon)}\int_{0}^{\infty}\!\! \!ds\ s^{\epsilon-1}K(s,f,
\mathcal{M})\,.
\end{equation}
As is customary, below we use $f=1$. However this should be understood as taking the
  limit $f\to1$ at the end of the computation to ensure all expressions in
  intermediate steps are well defined.

Formally $W$ is given by the divergent expression
\begin{align*}
W= -\int_{0}^{\infty} ds \frac{1}{s} K (s,1,\mathcal{M})
\end{align*}
The regularized version, $W_{\epsilon}$, is defined by shifting the power of $s$
\begin{align}
\label{eq:Weps}
W_{\epsilon}=-\tilde{\mu}^{2\epsilon}\int_{0}^{\infty} ds \frac{1}{s^{1-\epsilon}} K (s,1,\mathcal{M}) =-\tilde{\mu}^{2\epsilon}\Gamma(\epsilon)\zeta(\epsilon, 1,\mathcal{M})
\end{align}
where the parameter $\tilde{\mu}$ with length dimension $-1$ is
introduced so that $W_\epsilon$ remains adimensional. In this context, the parameter $\epsilon$ behaves like a regulator, the
divergences re-appearing  as $\epsilon\to 0$.
In this limit
\begin{align*}
W_{\epsilon}=-\left(\frac{1}{\epsilon}-\gamma_{E}+\ln(\tilde{\mu}^{2})\right)\zeta(0,1,\mathcal{M}) -\zeta^{\prime}(0,1,\mathcal{M})+ O(\epsilon)\,,
\end{align*}
so that subtracting  the $\frac{1}{\epsilon}$ term gives the
renormalized effective action 
\begin{align}
\label{eq:Wren}
 W^{\text{ren}}=
  -\zeta^{\prime}(0,1,\mathcal{M})-\ln(\mu^{2})\zeta(0,1,\mathcal{M})\,.
\end{align}
where $\mu^{2}=\tilde{\mu}^{2}e^{-\gamma_{E}}$ and \bg{$\gamma_E$ is the Euler constant}. On a compact manifold
$\zeta(\epsilon, 1,\mathcal{M})$ is finite as $\epsilon\to0$ and the
renormalized effective action given by \eqref{eq:Wren} is finite, as
it should. For non-compact manifolds the standard procedure for
computing a renormalized effective action is to subtract a reference
action that does not modify the physics. One may, for example, define
$W=\ln(\det(\mathcal{M})/\det(\mathcal{M}_0))$, where the
operator  $\mathcal{M}_0$ is defined on a trivial (flat)
background. This amounts to replacing $K(s,1,\mathcal{M})\to
K(s,1,\mathcal{M})-K(s,1,\mathcal{M}_0)$ in Eq.~\eqref{eq:Weps} and
correspondingly $\zeta(\epsilon, 1,\mathcal{M})\to \zeta(\epsilon,
1,\mathcal{M})-\zeta(\epsilon, 1,\mathcal{M}_0)$. The expression for
$W^{\text{ren}}$ in \eqref{eq:Wren} remains valid if it is understood
that this subtraction is made before the $\epsilon\to 0$ limit is
taken.

Classical symmetry under Weyl variations (both in the  
relativistic case and   the anisotropic one) guarantees $\mathcal{M}$
transforms homogeneously, {\it i.e.,} $\delta_\sigma\mathcal{M}= -\Delta\sigma \mathcal{M}$ under
$\delta_\sigma g_{\mu\nu}= {2\sigma}g_{\mu\nu}$
where $\Delta$ is the scaling dimension of $\mathcal{M}$. Hence, we have 
\begin{align}
\delta_\sigma \zeta (\epsilon,1,\mathcal{M}) = \bg{-\epsilon\Tr_{L^{2}} \left(
  \delta\mathcal{M}\mathcal{M}^{-\epsilon-1}\right)} = \Delta \epsilon \zeta (\epsilon, \sigma, \mathcal{M})\,.
\end{align}
Consequently, the anomalous variation of $W$ is given by 
\begin{align}
 \delta_\sigma W^{ren} = -\Delta\zeta(0,\sigma,\mathcal{M})\,.
\end{align}
In the relativistic case, using the fact that 
\begin{align}
\delta_{\sigma}W = \frac12\int d^{d+1}x \sqrt{g} T_{\mu\nu}\delta g^{\mu\nu}= - \int d^{d+1}x
  \sqrt{g} T^{\mu}{}_{\mu} \delta\sigma\,,
\end{align}
one has the trace anomaly equation
\begin{align} 
\mathcal{A}=-T^{\mu}{}_{\mu}=-\frac{1}{\sqrt{g}}\Delta\left(\frac{\delta\zeta(0,\sigma,\mathcal{M})}{\delta\sigma}\right)_{\sigma=0}\,.
\end{align}
In the non-relativistic case, the Weyl anisotropic scaling is given by $h_{\mu\nu}\to
e^{2\sigma}h_{\mu\nu}$ and $n_{\mu}\to e^{z\sigma}n_{\mu}$. We have
\begin{align}
\delta_{\sigma}W = \int d^{d+1}x \sqrt{g} \left(\frac{1}{2} T_{\mu\nu} \delta h^{\mu\nu} - \mathcal{E}_{\mu} \delta n^{\mu}\right)= \int d^{d+1}x \sqrt{g} \left(h^{\mu\nu}T_{\mu\nu}- zn^{\mu}\mathcal{E}_{\mu}\right) \delta\sigma
\end{align}
leading to
\begin{align}
\mathcal{A}=zn_{\mu}\mathcal{E}^{\mu}-h^{\mu\nu}T_{\mu\nu}=-\frac{1}{\sqrt{g}}{\Delta}\left(\frac{\delta\zeta(0,\sigma,\mathcal{M})}{\delta\sigma}\right)_{\sigma=0}\,.
\end{align}

\bg{One can evaluate \bg{$\left.\delta\zeta(0,\sigma,\mathcal{M})/\delta\sigma\right|_{\sigma=0}$} using the asymptotic form ($s\to0$) of \bg{the} heat kernel\bg{,} $K$. The asymptotic expansion depends on the operator $\mathcal{M}$ and its scaling dimension. Schematically, one has 
\bg{\begin{align}
\nonumber K(s,1,\mathcal{M})= \frac{1}{s^{d_{\mathcal{M}}}}\sum_{n=0}^{\infty}\lsp s^{\kappa(n)}\lsp \sqrt{g}\lsp a_{n},
\end{align}
where $\kappa(n)$ is a linear function of $n$. T}he singular \bg{pre-factor,
$\frac{1}{s^{d_{\mathcal{M}}}}$, is determined by the heat kernel in the background-free, flat space-time
limit while the expansion accounts for corrections from background fields or
  geometry}. The asymptotic expansion is guaranteed to exist if the heat kernel is well
behaved \bg{for $s> 0$ in the}  flat space-time limit, \bg{that is, if  $\sum_{i}e^{-s\lambda_i}$, with $\lambda_{i}$, the eigenvalues of the
operator $\mathcal{M}$,  is 
convergent. The convergence requires that  $\lambda_i$ have,} at worst, a power law growth and positive real
part~\citep{gilkey1980spectral}.

We are interested in operators $\mathcal{M}$ of generic form 
\begin{align}
\nonumber \mathcal{M}= 2 \imath m \partial_{t^{\prime}}- (-1)^{z/2}(\partial_{i}\partial_{i})^{z/2}\,,
\end{align}
\bg{for which} the  heat kernel 
has a small $s$ expansion of the following form
 \begin{align}
\label{eq:Gexp}
K(s,1,\mathcal{M})=
  \frac1{s^{1+d/z}}\sum_{n=0}^{\infty}\lsp s^{2n/z} \int d^{d+1}x\lsp \sqrt{g}\lsp a_{n}\,,
\end{align}
 where $d$ is number of spatial dimension and $z$ is dynamical exponent.\footnote{In next few sections, we explicitly find this asymptotic form for $z=2$ while the arbitrary $z$ case is handled separately in \ref{sec:gener}.}} Then the  zeta function is given by
\begin{align}
 \zeta(0,f,\mathcal{M})=\int d^{d+1}x \sqrt{g}\lsp f\lsp a_{(d+z)/2}\,,
\end{align}
so that  we arrive at an 
expression for the Weyl anomaly 
\begin{align}\label{anomalyheat}
\mathcal{A}= -{\Delta} \ a_{(d+z)/2}\,.
\end{align}

Hence, in order to determine the  Weyl anomaly, one has to calculate the coefficient
$a_{(d+z)/2}$ of the heat kernel expansion \eqref{eq:Gexp}.\footnote{Incidentally, this shows that
the anomaly is absent when $d+z$ is odd.} In
subsequent sections, we will find out a way to evaluate the  heat kernel in
flat space-time and then in curved space-time for a Schr\"odinger
invariant field theory. We will be doing this first without using
DLCQ/LCR, and then again with LCR (modified cousin of
DLCQ) using the prescription introduced above. 

\subsection{Heat Kernel in Flat Space-time}
\label{sec:heatKdircalc}
\subsubsection{Direct calculation (without use of  DLCQ)}
\label{sec:heatKdircalcsub}
The action for a free Galilean CFT on a flat space-time (which is in fact invariant under
the Schr\"odinger group) is given by
\begin{equation}
S= \int dt\ d^{d}x \phi^{\dagger}\left[2m\imath\partial_{t}+\nabla^{2}\right]\phi
\end{equation}
In order to improve convergence of the functional integral defining
the partition function we perform a continuation to imaginary time :
\begin{equation}
e^{\imath \int dt d^{d}x \phi^{\dagger}\left[2m\imath\partial_{t}+\nabla^{2}\right]\phi}\underset{t=-\imath\tau}{\mapsto}e^{- \int d\tau d^{d}x \phi^{\dagger}\left[2m\partial_{\tau}-\nabla^{2}\right]\phi}
\end{equation}
Hence, the Euclidean version of $\mathcal{M}=2m\imath\partial_{t}+\nabla^{2}$ is given by 
\begin{equation}
\mathcal{M}_{E}=2m\partial_{\tau}-\nabla^{2}\,,
\end{equation}
and it is this operator for which we will compute the heat kernel.
The prescription $t=-\imath\tau$ is equivalent to adding
$+\imath\epsilon$ to the propagator  in Minkowskian flat space. In
fact, the same $+\imath\epsilon$ prescription is obtained by deriving
the  non-relativistic propagator as the non-relativistic limit of the
relativistic propagator. 

The Heat kernel for $\mathcal{M}_E$ is a solution to the
equation\footnote{Even though $\mathcal{M}_{E}$ is not a hermitian
  operator, the heat kernel is well defined for any operator as long
  as $\text{Re}(\lambda_{k})>0$ where $\lambda_{k}$ are its eigenvalues. We
  explore this technical aspect in appendix.}
\begin{equation}\label{heatkerneleq}
(\partial_{s}+\mathcal{M}_E)\mathcal{G}=0\,,
\end{equation}
that is
\begin{align}\label{heatkerneleq1}
(\partial_{s}+2m\partial_{\tau_2}-\nabla^{2}_{x_2}) \mathcal{G}(s; (\vec{x}_2,\tau_2),(\vec{x}_1,\tau_1) )=0\,,
\end{align}
\bg{with boundary condition
$ \mathcal{G}(0; (\vec{x}_2,\tau_2), (\vec{x}_1,\tau_1))=
\delta(\tau_2-\tau_1)\delta^{d}(\vec{x}_2-\vec{x}_1)$}. Equation~\eqref{heatkerneleq1} is solved by 
\begin{equation}\label{heatkernelg00}
\mathcal{G}(s; (\vec{x}_2,\tau_2),(\vec{x}_1,\tau_1) ) =\delta\left(2ms - (\tau_2-\tau_1)\right)\frac{e^{-\frac{|\vec{x}_2-\vec{x}_1|^{2}}{4s}}}{(4\pi s)^{\frac{d}{2}}}
\end{equation}
Consequently, the Eulcidean two point correlator is given by
\begin{equation}
G((\vec{x}_2,\tau_2), (\vec{x}_1,\tau_1)) = \int_{0}^{\infty}\!\!\! ds\, \mathcal{G}(s) = \frac{\theta(\tau)}{2m}\frac{e^{-\frac{m|\vec{x}|^{2}}{2\tau}}}{(2\pi \frac{\tau}{m})^{\frac{d}{2}}}
\end{equation}
where $\tau=\tau_{2}-\tau_{1}$ and $\vec{x}=\vec{x}_{2}-\vec{x}_{1}$.
The same two point correlator can be obtained by Fourier transform from the Minkowski
momentum space propagator $G_{M}$, or its imaginary time version,
\begin{equation}
G_{M}(p,\omega) = \frac{\imath}{2m\omega -|\vec{p}|^{2}+i0^{+}} \underset{\underset{\omega=\imath\omega_{E}}{t=-\imath\tau}}{\mapsto} G=\frac{1}{2m\omega_{E} +\imath |\vec{p}|^{2}}
\end{equation}

In the coincidence limit the heat kernel of~\eqref{heatkernelg00} contains a
  Dirac-delta factor, $\delta(ms)$. Since this non-analytic behavior is unfamiliar, it is
  useful to re-derive this result by directly computing the trace $K$, \bg{Eq.~\eqref{heatkerneloperator}}. One can conveniently choose
the test function $f=e^{-|\eta\omega|}$. Hence
\begin{align*}
K(s,f,\mathcal{M}_{E,g})=\text{Tr}\ \left(fe^{-s\mathcal{M}_{E,g}}\right)=\int \left(\frac{d^{d}k}{(2\pi)^{d}} e^{-sk^{2}}\right)\left(\int \frac{d\omega}{2\pi}  e^{-2m\imath s \omega-|\eta\omega|}\right)
\end{align*}
The integral over $k$ gives the factor of $1/s^{d/2}$,
while the integral over $\omega$ gives
\[\frac{1}{\pi}\frac{\eta}{4m^{2}s^{2}+\eta^{2}}\]
that tends to 
$\delta(2ms)$ as $\eta\to0$. Before taking the limit,  this factor gives a
well behaved function for which the Mellin transform that
defines $\zeta$, Eq.~\eqref{eq:mellintransf}, is well defined for $d/2<\text{Re}(\epsilon)<d/2+2$
and can be analytically continued to $\epsilon=0$.

One may be concerned that the derivation above is only formal as it does not involve an elliptic operator.  This is easily remedied by considering the elliptic operator\footnote{The choice of regulator is suggested
  naturally, as it can ultimately be linked to the Minkowski form of the propagator 
\(
G= \frac{\imath}{2m\omega-k^{2}+\imath|\eta \omega|} \to \frac{\imath}{2m\omega-k^{2}+\imath 0^{+}}
\)
}  $\mathcal{M}^{\prime}=\imath \eta \sqrt{-\partial_{t}^{2}} + \imath (2m) \partial_{t} +\nabla^{2}$. Its  spectrum, $\left(2m\omega-k^2+\imath \eta|\omega|\right)$, tends to that of the Minkowskian Schr\"odinger operator $\mathcal{M}$ as $\eta\to0$. Consequently, the spectrum for the Euclidean avatar\footnote{Alternatively, one can think of introducing the regulator, only after going over to the Euclidean version. The unregulated Euclidean operator, $\mathcal{M}_{E,g}=2m\partial_{\tau}-\nabla^{2}$ is regulated to $\mathcal{M}^{\prime}_{E,g}=2m\partial_{\tau}-\nabla^2+\eta\sqrt{-\partial^{2}_{\tau}}$.} ($\mathcal{M}^{\prime}_{E,g}$) of $\mathcal{M}^{\prime}$ becomes $\left(k^{2}+2m\imath\omega+|\eta\omega|\right)$ and the heat kernel for that operator is given by 
\begin{align*}
K(s,1,\mathcal{M}^{\prime}_{E,g})=\text{Tr}\ \left(e^{-s\mathcal{M}^\prime_{E,g}}\right)=\int \left(\frac{d^{d}k}{(2\pi)^{d}} e^{-sk^{2}}\right)\left(\int \frac{d\omega}{2\pi}  e^{-2m\imath s \omega-s|\eta\omega|}\right)
\end{align*}
The integral over $k$ gives the factor of $1/s^{d/2}$ as before,
while the integral over $\omega$ gives
\[\frac{1}{\pi s}\left(\frac{\eta}{4m^{2}+\eta^{2}}\right)\]
that tends to 
$\frac{1}{s}\delta(2m)$ as $\eta\to0$. As we will see later, Light Cone Reduction technique indeed reproduces this factor of $\delta(2m)$.
 
\subsubsection{Derivation using LCR}
In Euclidean, flat $d+2$ dimensional space-time,  the heat kernel $\mathcal{G}_{R,E}$ of a relativistic scalar field at free fixed point is given by \citep{Mukhanov:2007zz}
\begin{align}
\label{heatkernelrel}
\mathcal{G}_{R, E}(s;x_{2}^{M},x_{1}^{M}) =\frac{1}{(4\pi s)^{d/2+1}}e^{-\frac{(x_{1}-x_{2})^2}{4s}}
\end{align}
where the superscript reminds us that this is the relativistic case
and
$(x_{1}-x_{2})^2=(x_{1}^{M}-x^{M}_{2})(x^{N}_{1}-x^{N}_{2})\delta_{MN}$.

In preparation for using LCR, we rewrite the expression
\eqref{heatkernelrel} by first reverting to Minkowski space, $t=-\imath x^0$,  and then switching to
  light-cone coordinates.\footnote{Recall,  in the parent theory
  $x^{\pm}=\frac{1}{\sqrt{2}}(x^1 \pm t)$. Note that we are using a
  non-standard sign convention in the definition of $x^-$.} Using
$x^{\pm}=x^{\pm}_{2}-x^{\pm}_{1}$ we have:
\begin{equation}
 \mathcal{G}_{R, M} (s; (x^{+}_{2}, x^{-}_{2},\vec{x}_{2}), (x^{+}_{1}, x^{-}_{1}, \vec{x}_{1})) = \frac{1}{(4\pi s)^{d/2+1}}e^{-\frac{x^{+}x^{-}}{2s}-\frac{|\vec{x}|^2}{4s}} 
\end{equation}
where $\mathcal{G}_{R, M}$ is the heat kernel in Minkowski space. Now,
in the reduced theory, the co-ordinate $x^{+}$ becomes the time
coordinate $t$. Going to imaginary time, $t\to \tau = \imath t$, and
Fourier transforming we obtain the heat kernel $\mathcal{G}_{g,E}$ for
the Galilean invariant theory in Euclidean space:
\begin{align}
\mathcal{G}_{g,E}(s;(\vec{x}_2,\tau_2), (\vec{x}_1,\tau_1)) )&= \int_{-\infty}^{\infty } \frac{1}{(4\pi
                                                             s)^{d/2+1}}e^{\frac{\imath\tau
                                                             x^{-}}{2s}-\frac{|\vec{x}|^2}{4s}}
                                                             e^{-\imath m x^{-}} dx^{-}\nonumber\\ 
\label{heatkernelg10}
&= 2\pi \delta\left(\frac{\tau}{2s}-m\right)\frac{1}{(4\pi s)^{d/2+1}}e^{-\frac{|\vec{x}|^2}{4s}}
\end{align}
where $\tau=\tau_{2}-\tau_{1}$, in \bg{detailed} agreement with Eq.~\eqref{heatkernelg00}. For later use we note that in the coincidence limit we have 
\begin{align}\label{heatkernelg11}
\mathcal{G}_{g,E}((\vec{x},\tau), (\vec{x},\tau)) )=\frac{2\pi\delta(m)}{(4\pi s)^{d/2+1}}.
\end{align}
It is interesting to note that LCR directly gives $\sim \delta(m)/s^{d/2+1}$ while the direct computations gives $\sim\delta (ms)/s^{d/2}$. Our main result, below, follows from the
coincidence limit of the heat kernel expansion in Eq.~\eqref{heatkernelgcurved}, which is useful only for $s\ne0$,
since it is used to extract the coefficients of powers of $s$ in the
expansion. The limiting behavior as $s\to0$ of the function
$\mathcal{G}_{g,E}$ is a delta function enforcing coincidence of the
points, by construction (and this is why $a_0=1$ at coincidence), and
therefore the behavior as $s\to0$ is correct but of no significance.

The spectral dimension of the operator $\mathcal{M}_{E}$ is given by
\begin{equation}
d_{\mathcal{M}}= -\frac{d\ln(K)}{d\ln(s)} =\frac{d}{2}+1
\end{equation}
which explains why there can not be any trace anomaly when the spatial
dimension $d$ is odd. This has to be contrasted with the relativistic
case where the spectral dimension of the laplacian operator is
given by $\frac{d+1}{2}$, so that in the relativistic case the anomaly is only present when the spatial dimension $d$ is odd.


\subsection{Heat Kernel in Curved spacetime}
\label{sec:heatKcurv}
Now that we know that LCR works in flat space-time, we can go ahead
and implement it in curved space-time exploiting the known fact that
for relativistic field theories coupled to a curved geometry, the heat
kernel can be obtained as an asymptotic series. The method is
explained in, {\it e.g.,} Refs.~\citep{Jack:1983sk,Mukhanov:2007zz,Grinstein:2015ina}. 

The method,  first worked out by
DeWitt~\cite{DeWitt:1965jb},  starts with an Ansatz
for the form of the heat kernel taking a cue from the form of the
solution in flat space-time for the heat equation. For small enough $s$ the Ansatz for the heat kernel, corresponding to a relativistic theory in $d+2$ dimensions, reads:
\begin{equation}
\mathcal{G}_{R,E}(x_2,x_1; s) = \frac{\Delta_{\text{VM}}^{1/2}(x_2,x_1)}
{(4\pi s)^{d/2+1}}e^{-\sigma (x_2,x_1)/2s} \sum_{n=0}^\infty
a_{n} (x_2,x_1) \lsp s^n\,,\qquad
a_{0}(x_1,x_2)=1
\end{equation}
with $a_{n}(x_2,x_1)$ the so-called Seeley--DeWitt
coefficients and where $\sigma(x_2,x_1)$ is the biscalar distance-squared
measure (also known as the geodetic interval, as named by DeWitt), defined by
\begin{equation}\label{eq:geodeticinterval}
\sigma(x_2,x_1) = \tfrac{1}{2} \left( \int_0^1 d\lambda \,
\sqrt{G_{MN} \frac{dy^{M}}{d\lambda}\lsp
\frac{dy^{N}}{d\lambda}} \, \right)^2 \,,\qquad y(0)=x_1\,,\,\,y(1)=x_2\,,
\end{equation}
with $y(\lambda)$ a geodesic. The bi-function
$\Delta_{\text{VM}}(x_2,x_1)$ is called the van Vleck-Morette
determinant; this biscalar describes the
spreading of geodesics from a point and is defined by
\begin{align}\label{eq:vandet}
\Delta_{\text{VM}} (x_2,x_1) =G(x_2)^{-1/2}\lsp G(x_1)^{-1/2}
\:\det\!\left(-\frac{\partial^2}{\partial x_{2}^{M}
\partial x_{1}^{N^{\prime}}} \sigma (x_2,x_1) \right). 
\end{align}
where $G$ is the negative of determinant of metric $G_{MN}$.


\bg{Now, to implement LCR, recall that a Schr\"odinger invariant theory
coupled to a generic curved NC structure is obtained by reducing  from  the $d+2$ dimensional metric $G_{MN}$ in
Eq.~\eqref{eq:backgroundmetric}. \bg{In taking the coincident limit we must keep $x^{-}_{1}$ and
  $x^{-}_{2}$ arbitrary in order to Fourier transform with respect to $x^{-}$ per the
  prescription~\eqref{eq:prescription}.  Therefore, we  work in the coincident limit where
  $x^{\mu}_{1}=x^{\mu}_{2}$, with $\mu=+,1,2,\cdots, d$. } Now, since $x^{-}$ is a null direction,
in \bg{this} limit we have
\bg{$\sigma((x^{-}_{1},x^\mu),(x^{-}_2,x^\mu))=0$ or
  $\left[\sigma\right]=0$ for brevity}. Furthermore, null isometry guarantees that metric components
are independent of $x^{-}$ and so are \bg{$[a_{n}]$ and $[\Delta_{VM}]$.} Thus the coincident limit is equivalent to the coincident limit of the parent
theory, hence \bg{$\big[\Delta_{VM}\big]=1$}. We refer to appendix~\ref{riemann} for \bg{details}.}

\bg{Thus, in the coincidence limit, we have the following expression for the heat kernel corresponding to the reduced theory: 
\begin{equation}
\label{heatkernelgcurved}
\mathcal{G}_{g,E}(s; (\tau,\vec{x}), (\tau,\vec{x}) ) =\frac{2\pi\delta(m)}{(4\pi s)^{d/2+1}}\sum_{n=0}^\infty
a_{n} ((\tau,\vec{x}),(\tau,\vec{x})) \lsp s^n\,,\qquad
a_{0}((\tau_1,\vec{x}_1),(\tau_2,\vec{x}_2))=1
\end{equation}
where to define $\tau$, we have proceeded just as in flat
space: first revert to a Minkowski metric, then switch to light cone coordinates, and finally go over to imaginary $x^+$ time, $\tau$.}
Subsequently, using Eq.~\eqref{anomalyheat} the anomaly is given by
\begin{align}\label{an}
\mathcal{A}^{G}_{d+1}=-4\pi \delta(m) \frac{a_{d/2+1} }{(4\pi)^{d/2+1}} .
\end{align}
From Eq.~\eqref{heatkernelgcurved} it is clear that only the zero
mode of $P_-$ can contribute to the anomaly; the anomaly vanishes for fields with non-zero
$U(1)$ charge. We already know that the anomaly for the relativistic complex scalar case is given by
\begin{align}
\mathcal{A}^{R}_{d+2}=-\frac{2a_{d/2+1} }{(4\pi)^{d/2+1}}\,.
\end{align}
Thereby we establish the result advertised in the introduction, giving the Weyl anomaly of
a $d+1$ dimensional Schr\"odinger invariant field theory of a single complex scalar field
carrying charge $m$ under $U(1)$ symmetry), $\mathcal{A}^{G}_{d+1}$, in terms of the
anomaly in the relativistic theory in $d+2$ dimensions, $\mathcal{A}^{R}_{d+2}$:
\begin{align}
\mathcal{A}^{G}_{d+1}=2\pi \delta(m) \mathcal{A}^{R}_{d+2}\,,
\end{align}
computed on the class of metrics given in Eq.~\eqref{eq:backgroundmetric}. 

\bg{At this point, we pause to remark on \bg{the interpretation of the $\delta(m)$
    factor}. \bg{While it trivially}  shows that the anomaly is
  absent for $m\neq 0$ \bg{, t}he interpretation becomes subtle when $m=0$. \bg{The apparent divergence in the anomaly} is just
  an artifact of \bg{the} usual zero mode problem associated with null reduction. \bg{A} similar
  issue has been pointed out in \citep{Jensen:2014aia} \bg{in} reference
  to~\citep{Maldacena:2008wh,Adams:2008wt}. The reduced theory in \bg{the} $m\to0$ limit becomes
  infrared divergent\bg{;} the fields become non-dynamical in that limit. The infrared
  divergence is also evident \bg{from E}q.~\eqref{eq:prescription}. One \bg{may}
  further \bg{understand} the presence of $\delta(m)$ by \bg{letting $m$ be a continuous parameter} and considering a continuous set of fields $\phi_m$,
  \bg{of} charge $m$. \bg{T}he anomaly \bg{arising} from the continuous set of fields is given by \bg{summing over their
  contributions:}
\begin{align}
\nonumber  \frac{1}{2\pi}\lsp \int\lsp dm\lsp \mathcal{A}^{G}_{d+1}= \mathcal{A}^{R}_{d+2}\lsp \int\lsp dm\lsp \delta(m)=\mathcal{A}^{R}_{d+2}
\end{align}
The right hand side is exactly what we expect since allowing the parameter $m$ to continuously vary
restores the Lorentz invariance\bg{: consulting Eq.~\eqref{actiong} we see that this continuous sum corresponds to restoring
  the relativistic theory of Eq.~\eqref{actionr}.}}

That the constant of proportionality relating $\mathcal{A}^{R}_{d+2}$ to $\mathcal{A}^{G}_{d+1}$
vanishes for $m\neq 0$ can be verified by an all-orders computation of $\mathcal{A}^{G}_{d+1}$, to
which we now turn our attention. 
\section{Perturbative proof of Vanishing anomaly}
\label{perturbation}
The fact that the anomaly  vanishes for  non-vanishing $m$ can be shown
perturbatively taking the background to be slightly
curved. In flat space-time, wavefunction renormalization
and coupling constant renormalization are sufficient to render a
quantum field theory finite. Defining composite operators
requires further renormalization. Therefore, when the model is placed
on a curved background additional short distance divergences appear
since the background metric can act as a source of operator
insertions. To cure these divergences, new counter-terms are required
that may break scaling symmetry even at a fixed point of the
renormalization group flow.  In this section, we will treat the
background metric as a small perturbation of a flat metric so that we
compute in a field theory in flat space-time with the effect of
curvature appearing as operator insertions of the perturbation
$h_{\mu\nu}=g_{\mu\nu}-\eta_{\mu\nu}$.  To be specific, we will look
at the vacuum bubble diagrams with external metric insertions. It
turns out that all of these Feynman diagrams vanish at all orders of
perturbation theory, leading to a vanishing anomaly.  In fact, \bg{we
will show} that these anomalies 
vanish even away  from the fixed point as long as the theory 
satisfies some nice properties.

Suppose we have a rotationally invariant field theory such that:
\begin{enumerate}
\item The theory includes only rotationally invariant (``scalar'')  fields. 
\item At free fixed point, the theory admits an $U(1)$ symmetry under
  which the  scalar fields are
  charged.
\item \label{thm:prop} \bg{The free propagator is of the form
  $\frac{\imath}{2m\omega - f(|\vec{k}|)+\imath\epsilon}$, where, generically,
  $f(|\vec{k}|)=|\vec{k}|^{z}$.}
\item \label{thm:int} The interactions are perturbations about the free fixed point by
  operators of the form $g(\phi,\phi^*)|\phi|^{2}$, where $g$ is a
  polynomial of the scalar field $\phi$.
\end{enumerate}

An elementary argument presented below shows that, under these conditions, all the vacuum bubble
diagrams vanish to all orders in perturbation theory.

Before showing this, a few comments are in order. First, the argument is valid in any
number of spatial dimensions.  Second, assumption~\ref{thm:int} precludes terms like
$\phi^{4}+(\phi^{*})^{4}$ or $K\phi^{2}$ in the Lagrangian. To be precise,
$F(\phi)+\text{h.c.}$ can evade this theorem for any holomorphic function $F$ of
$\phi$. This is because assumption~\ref{thm:int} implies that each vertex of the Feynman
diagrams of the theory has at least one incoming scalar field into it and one outgoing
scalar field line from it; having both incoming and outgoing lines at each vertex is at
the heart of this result. Thirdly, it should be understood that all interactions that can
be generated via renormalization, that is, not symmetry protected, are to be included. For
example, were we to consider a single scalar field with only the interaction
$\phi^{3}\phi^{*}+\text{h.c.}$, the interactions $\phi^{4}+(\phi^{*})^{4}$ and
$(\phi\phi^{*})^{2}$ will be generated along the RG flow. Nonetheless, $U(1)$ symmetry
will always prohibit a holomorphic interaction $F(\phi)+h.c$. Lastly,
assumption~\ref{thm:prop} can be relaxed to include a large class of functions
$f(|\vec{k}|^{2})$; this means one can recast this result in terms of perturbation theory
along the RG-flow rather than about fixed points.


To prove this claim, notice first that a vacuum diagram is a connected graph without
external legs (hanging edges). Moreover, since we are considering a complex scalar field,
the vertices are connected by directed line segments. These directed segments form
directed closed paths. To see this, recall that by assumption each vertex has at least one
ingoing and one outgoing path. Starting from any vertex, we have at least one outgoing
path. Any one of these paths must have a second vertex at its opposite end, since by assumption there are not hanging edges.  Take any one outgoing path  and follow it to the next
vertex. Now, at this second vertex repeat this argument: follow the outward
path to a third vertex. And so on. Since a finite graph has a finite number of vertices, at some point in the process we have to come back to a vertex we have already visited. For example, assume that we first revisit  the  $i$-th
vertex. This means that starting from the vertex
$i$ we have a directed path which loops back to the $i$-th vertex itself. The simplest example is that of a path starting and ending on the first vertex, corresponding to a self contraction of the elementary field in the operator insertion.

Let us call this directed loop $\Gamma$. We use the freedom in the choice of loop energy and momentum in the evaluation of  the Feynman diagram  to assign a
loop energy $\omega$ in a way such that $\omega$ loops around $\Gamma$.  In performing the integral over  $\omega$  it suffices to consider the $\Gamma$ subdiagram only. The resulting integration is of the form:
\begin{equation}
\int d\omega\, P(\omega,\vec k, \{\omega_n,\vec{ k}_n\})\prod_{n\in\Gamma}\frac{1}{(\omega+\omega_n-f(|\vec k+ \vec{k}_n|)/2m+i\epsilon)} 
\end{equation}
where the product is over all vertices in $\Gamma$ and correspondingly over all line
segments in $\Gamma$ out of these vertices. Energy $\omega_n$ and momentum $\vec k_n$
enter $\Gamma$ at the vertex $n$.  The factor $P(\omega,\vec k, \{\omega_n,\vec{ k}_n\})$
is polynomial in momentum and energy and may arise if there are derivative interactions. Note that every propagator factor has the same sign $i\epsilon$ prescription, that is, all poles in complex-$\omega$ lie in the lower half plane (have negative imaginary part). The integral  over the real $\omega$ axis can be turned into an integral over a closed contour in the complex plane, by closing the contour on an infinite radius semicircle on the upper half plane, using the fact that for two or more propagators the integral over the semicircle at infinity vanishes. Then Cauchy's theorem gives that the integral over the closed contour vanishes as there are no poles inside the contour. 

This proves the claim, except for the singular case of a self-contraction, that is, a
propagator from one vertex to itself. Self contractions can be removed by normal ordering,
again giving a vanishing result. For an alternative way of seeing this note that this
integral is independent of external momentum and energy, and is formally divergent in the
ultraviolet (as $|\omega|\to\infty$). The integral results in a constant (independent of
external momentum and energy) that must be subtracted to render it finite, and can be chosen to be subtracted
completely, to give a vanishing result. 

The computation in the case of anti-commuting fields differs only in that a factor of $-1$ is
introduced for each closed fermionic loop. Hence the claim applies equally to the case of
anti-commuting scalar fields. 

We now return to the derivation of our main result, Eq.~\eqref{mainresult}. The conditions
above are satisfied for the theories considered in Sec.~\ref{sec:heatKcurv}, namely, free
theories of complex scalars, with the free propagator given by
$\frac{\imath}{2m\omega - |\vec{k}|^{2}+\imath 0^{+}}$. Recall that we are to put the
theory on a curved background which is assumed to be a small perturbation from flat
background. The perturbations act as insertions on vacuum bubble diagrams, but since they
preserve the $U(1)$ symmetry the model still satisfies the assumptions above. Hence all the bubble
diagrams vanish, and we conclude there are no divergences coming from metric insertion on
bubble diagrams. Consequently, there is no scale anomaly. We emphasize that the absence of
the Weyl anomaly is valid in all orders of perturbation in both the coupling and the
metric. The result holds true even if we make the couplings to be space-time dependent so
that every coupling insertion injects additional momentum and energy to the bubble
diagram. Physically,  the anomaly vanishes  because  the absence of antiparticles in
non-relativistic field theories and the conservation of $U(1)$ charge forbid pair creation,
necessary for  vacuum fluctuations that may give rise to the anomaly.

This perturbative proof holds for theories which need not be Galilean invariant, and the
question arises as to whether one may use LCR to make statements about anomalies for
theories with kinetic term involving one time derivative and $z\neq 2$. We will take up
this task in following section, starting by giving the promised proof of
our prescription in Eq.~\eqref{eq:prescription}.
%

\bg{We remark that perturbative proof works for $m\neq 0$. For $m=0$,
  the integrand becomes independent of $\omega$, \bg{and}
  one can not \bg{perform} the contour integral to argue the
  diagrams vanish. In fact,  \bg{the integral over
  $\omega$ is divergent, as expected from our earlier expectation that
  at $m=0$ one encounters IR divergences}.  One way to see the
  presence of $\delta(m)$\bg{, as explained earlier,} is to take a continuous set of fields
  $\phi_m$, labelled by continuous parameter $m$. \bg{If} we
  exchange the sum over \bg{(1-loop)} bubble diagrams and \bg{the} integral over $m$, then
  each of the propagator can be thought of as a relativistic
  propagator with $m$, playing the role of $p_-$. Thus the
  whole calculation formally becomes \bg{that of the}
  relativistic anomaly.}

\bg{One can verify our results by explicit calcualation in specific
  cases. In a slightly curved space-time, one can
  treat the deviation from flatness  as background field sources. This also serves the purpose of checking that the
  $\eta$-regularization is appropriate, obtaining the
  anomaly as a function of $\eta$.  Since, as $\eta\to0$, for $m\neq 0$,
  the flat space heat kernel vanishes, one expects the anomaly to be
  vanishing. In fact, one can check that  a $\delta(m)$ is
  recovered as $\eta\to0$. We refer to the App.~\ref{hketa} for an
  explicit calculation; it verifies our results in detail, and shows
  the vanishing anomaly regardless of the order of 
   limits $\eta\to0$ and $m\to0$.}


\section{Modified LCR and Generalisation}\label{generalisation}
\subsection{Proving the heat kernel prescription}
In this subsection we will explain why our proposed method to determine the heat kernel
for Schr\"odinger field theory ($z=2$) worked in a perfect manner, as evidenced by the
agreement between Eqs.~\eqref{heatkernelg00} and~\eqref{heatkernelg10}. We will see
that one can use LCR to relate the heat kernel of a theory living in $d+1$ dimensions
with that of a parent theory living in $d+2$ dimensions, as long as the parent theory has
$SO(1,1)$ invariance.\footnote{One may as well assume that both parent and reduced
  theories have, in addition, $SO(d)$ rotational symmetry.}  Furthermore, if the parent theory has a dynamical scaling exponent
given by $z$, then the theory living in $d+1$ dimension has $2z$ as its dynamical
exponent. We will make these statements precise in what follows.

Suppose the operator $D$ defined in $d+2$ dimensional space-time  is
diagonal in the eigenbasis of $P_-$, the conjugate momenta to $x^{-}$:
\begin{align}
\langle x^{+}_{2}, x^{i}_{2}, m_{2}| D| x^{+}_{1},x^{i}_{1},m_{1}\rangle = \langle x^{+}_{2}, x^{i}_{2}| D_{m_2}| x^{+}_{1},x^{i}_{1}\rangle \delta(m_{2}-m_{1})\,,
\end{align}
where $m_{1,2}$ label the eigenvalues of $P_-$. The example  worked out in Sec.~\ref{sec:heatKdircalc}
had  $D=\mathcal{M}$, and it does satisfy this requirement.
It follows that 
\begin{align}
\nonumber \langle x^{+}_{2}, x^{i}_{2}, x^{-}_{2}| e^{-sD}| x^{+}_{1},x^{i}_{1},x^{-}_{1}\rangle &=&&\frac{1}{2\pi} \int dm_{1}\ dm_{2} e^{-\imath m_{1}x^{-}_{1}+\imath m_{2}x^{-}_{2}}\langle x^{+}_{2}, x^{i}_{2}, m_{2}| e^{-sD}| x^{+}_{1},x^{i}_{1},m_{1}\rangle\\ &=&& \frac{1}{2\pi} \int dm_{1} e^{\imath m_1 x^{-}_{12}}\langle x^{+}_{2}, x^{i}_{2}| e^{-sD_{m_1}}| x^{+}_{1},x^{i}_{1}\rangle\,,
\end{align}
from which we obtain
\begin{align}
\label{eq:prescription2}
\langle x^{+}_{2}, x^{i}_{2}| e^{-sD_{m}}| x^{+}_{1},x^{i}_{1}\rangle = \int dx^{-} e^{-\imath m x^{-}_{12}} \langle x^{+}_{2}, x^{i}_{2}, x^{-}_{2}| e^{-sD}| x^{+}_{1},x^{i}_{1},x^{-}_{1}\rangle\,.
\end{align}
This is precisely the prescription we gave in Eq.~\eqref{eq:prescription}.

\subsection{Generalisation}
\label{sec:gener}
Since the LCR (or DLCQ) trick requires null cone reduction, it may seem necessary that
the parent theory have $SO(d+1,1)$ symmetry, and that this  will result necessarily in a Galilean invariant reduced
theory, that is,  with $z=2$. This is not quite right: one may relax the condition of $SO(d+1,1)$
symmetry and obtain reduced theories with $z\ne 2$. The key observation is that for null
cone reduction only two null coordinates are needed, with the rest of the coordinates
playing no role. Hence, we consider null cone reduction of a $d+2$ dimensional theory
which enjoys $SO(1,1) \times SO(d)$ symmetry.  The reduced theory will be a $d+1$
dimensional theory with $SO(d)$ rotational symmetry and a residual $U(1)$ symmetry that
arises from the null reduction. The point is that the theory can enjoy anisotropic scaling
symmetry. Consider, for example, the following class of operators
\begin{align}
\label{eq:Mrd+2}
\mathcal{M}_{rc; d+2} = \left(-\partial_{t}^{2}+\partial_{x}^{2}\right) - (-1)^{z/2} (\partial_{i}\partial^{i})^{z/2}\,,
\end{align}
where $t=x^0$ and $x=x^{d+1}$ and for the reminder of this section there is an implicit
sum over repeated latin indices, over the range $i=1,\ldots,d$. These operators transform
homogeneously under
\begin{align}
x^{i}\to \lambda x^{i},\qquad t\to \lambda^{z/2}t\qquad\text{and}\qquad  x\to \lambda^{z/2}x\,. 
\end{align} 
Introducing null coordinates as before, $x^{\pm}=\frac{1}{\sqrt{2}}(x\pm t)$, 
null reduction of this operator yields 
\begin{align}
\mathcal{M}_{gc; d+1}= 2 \imath m \partial_{t^{\prime}}- (-1)^{z/2}(\partial_{i}\partial_{i})^{z/2}\,,
\end{align}
where $t^{\prime}=x^{+}$ is the time coordinate of the reduced theory. From the dispersion
relation of the reduced theory, $2m\omega = |\vec k|^{z}$, we read off that the dynamical
exponent is $z$. \bg{Here we are interested in even $z$ to \bg{insure} that the operator $\mathcal{M}_{gc; d+1}$ is local.} For $z=2$, we recover the case discussed in earlier sections with the
parent theory being Lorentz invariant and the reduced theory being Schr\"odinger
invariant. 

Following the prescription~\eqref{eq:prescription2}, we can relate the matrix element of the heat
kernel operator for $\mathcal{M}_{r;d+2}$ to that of $\mathcal{M}_{g;d+1}$,
via\footnote{Provided these heat kernels are well defined. We postpone this technical
  aspect to the appendix. }
\begin{align}
\label{mastereq}
\mathcal{G}_{\mathcal{M}_{gc;d+1}}= \int_{-\infty}^{\infty} dx^{-}\ e^{-\imath m x^{-}} \langle x^{-}_{0}+x^{-}|\mathcal{G}_{\mathcal{M}_{rc;d+2}}|x^{-}_{0}\rangle\,.
\end{align}
This should be viewed as an operator relation: thinking of the basis on which the operator
$\mathcal{G}_{\mathcal{M}_{r;d+2}}$ acts as given by the tensor product of $|x^{+}\rangle$,  $|x^{-}\rangle$ and  $|x^{i}\rangle$ for
$i=1,2,\ldots,d$,  then 
$ \langle x^{-}_{0}+x^{-}|\mathcal{G}_{\mathcal{M}_{r;d+2}}|x^{-}_{0}\rangle $ is an
operator acting on the complement of the space spanned by $|x^{-}\rangle$. Taking  the trace on both  sides of Eq.~\eqref{mastereq}, we obtain the heat kernel of the reduced theory:
\begin{align}\label{mastereq1}
K_{\mathcal{M}_{gc;d+1}}= \int_{-\infty}^{\infty} dx^{-}\ e^{-\imath m x^{-}}\ \Tr_{x^{+},x^{i}}  \langle x^{-}_{0}+x^{-}|\mathcal{G}_{\mathcal{M}_{rc;d+2}}|x^{-}_{0}\rangle
\end{align}

Equations \eqref{mastereq} or \eqref{mastereq1} are useful in practice only when we know
either left or right hand sides by some other means. Hence, the next meaningful question
to be asked is whether we can calculate $\mathcal{G}_{\mathcal{M}_{r}}$ explicitly for a curved
space-time for any $z$. The case for $z=2$, that in which the parent theory is
relativistic and the reduced theory is Schr\"odinger invariant, is well known and was
presented in Sec.~\ref{sec:heatKdircalc}. For generic $z$, the answer is yes to some
extent. We will find a closed form expression when the slice of constant $(t,x)$ in
space-time is described by a metric that does not depend on $t$ or $x$:
\begin{equation}
\label{eq:metricslice}
ds^{2} = -dt^{2}+(dx)^{2}+h_{ij}(x^{i})dx^{i}dx^{j}
\end{equation}
With this choice, the heat kernel equation for the curved background version of the
operator $\mathcal{M}_{rc;d+2}$ of Eq.~\eqref{eq:Mrd+2} admits a solution by separation of
variables, into the product of the relativistic heat kernel in $1+1$ dimensions and the
heat kernel for an operator acting only on the $d$-dimensional slice  \citep{Baggio:2011ha}. Specifically, we
consider operators
\begin{align}
\label{curvedoperator}
\mathcal{M}_{rc;d+2}= \nabla^{2}_{t,x}-D^{z/2}
\end{align}
where $\nabla^{2}_{t,x}=(-\partial_{t}^{2}+\partial_{x}^{2})$ and $D$ is a second order
scalar differential operator on the slice of constant $(t,x)$, {\it e.g.},
$D=-\nabla^{2}=-1/\sqrt{h}\,\partial_i\sqrt{h}h^{ij}\partial_j$.  With these choices,
\begin{equation}
\mathcal{G}_{\mathcal{M}_{rc;d+2}}=\mathcal{G}_{\nabla^{2}_{t,x}}\;\mathcal{G}_{D^{z/2}}\,.
\end{equation}
Gilkey has shown that the  heat kernel expansion for $D^k$ can be computed from that for $D$~\citep{gilkey1980spectral} \bg{for $k>0$}. The argument is based on the observation that  the $\zeta$-functions for the two operators are related:
\[
 \zeta (\epsilon , f,D^k ) = \Tr_{L^{2}} \left(f (D^k)^{-\epsilon}\right) =\Tr_{L^{2}} \left(f D^{-k \epsilon}\right) = \zeta (k \epsilon , f,D ) \,.
\]
Gilkey's result is as follows:  If $D$ has heat kernel expansion 
\begin{align}
K_D=\left(\frac{1}{\sqrt{4\pi}}\right)^{d}\sum_{n\geq 0} s^{n-\frac{d}{2}}a^{(d)}_{n}
\end{align}
then the heat kernel expansion of $D^k$ is 
\begin{align}
\nonumber K_{D^k}= \left(\frac{1}{\sqrt{4\pi}}\right)^{\!\!d}\sum_{n\geq 0} s^{\frac{2n-d}{2k}}\frac{\Gamma(\frac{d-2n}{2k})}{k\Gamma(\frac{d}{2}-n)}a^{(d)}_{n} &=\left(\frac{1}{\sqrt{4\pi}}\right)^{\!\!d}\sum_{\underset{2n\neq d (\text{mod\ } 2k)}{n\geq 0}} s^{\frac{2n-d}{2k}}\frac{\Gamma(\frac{d-2n}{2k})}{k\Gamma(\frac{d}{2}-n)}a^{(d)}_{n} \\
&+\left(\frac{1}{\sqrt{4\pi}}\right)^{\!\!d}\sum_{\underset{2n=d (\text{mod\ } 2k)}{n\geq 0}} s^{\frac{2n-d}{2k}}(-1)^{\frac{(2n-d)(1-k)}{2k}}a^{(d)}_{n}
\end{align}
%
%
Hence, $\mathcal{M}_{rc;d+2}=(-\partial_{t}^{2}+\partial_{x}^{2}) - (-\nabla^{2})^{z/2}$ has heat kernel expansion
\begin{align}
\label{eq:KMrcexp}
\langle x^{+}_{2},x^{-}_2, x^{i}| \mathcal{G}_{\mathcal{M}_{rc;d+2}}| x^{+}_{1},x^{-}_{1},x^{i}\rangle 
= \frac{e^{\frac{-x^{+}_{12}x^{-}_{12}}{2s}}}{4\pi s}\left(\frac{1}{\sqrt{4\pi}}\right)^{\!\!d}
\sum_{n\geq 0} s^{\frac{2n-d}{z}}\frac{\Gamma(\frac{d-2n}{z})}{\frac{z}2\Gamma(\frac{d}{2}-n)}a^{(d)}_{n}
\end{align}
where $x^{\pm}_{12}=x^{\pm}_{2}-x^{\pm}_{1}$ and $a^{(d)}_{n}$ are the well known
coefficients of the heat kernel expansion of~$ -\nabla^{2}$.

Now, the reduced theory lives on $d+1$ dimensional space-time with curved spatial slice, {\it i.e.}, the background metric is given by
\begin{align}
ds^{2}=-dt^{2}+h_{ij}dx^{i}dx^{j}\,,
\end{align}
where $i$ runs from $1$ to $d$. In order to extract the heat kernel of $\mathcal{M}_{gc;d+1}=2\imath m\partial_{t}+(-\nabla^2)^{z/2}$, we need partial tracing of heat kernel of $\mathcal{M}_{rc;d+2}$,
\begin{align}
\langle x^{-}_{0}+x^{-}| \Tr_{x^{+},x^{i}} \mathcal{G}_{\mathcal{M}_{rc;d+2}}|x^{-}_{0}\rangle 
= \left(\frac{1}{\sqrt{4\pi}}\right)^{\!\!d} \frac{1}{4\pi s}\sum_{n\geq 0} s^{\frac{2n-d}{z}}\frac{\Gamma(\frac{d-2n}{z})}{\frac{z}2\Gamma(\frac{d}{2}-n)}a^{(d)}_{n}\,,
\end{align}
leading to 
\begin{align}
\label{eq:KMgcexp}
K_{\mathcal{M}_{gc;d+1}}= 2\pi \delta(m)\frac{1}{4\pi s} \left(\frac{1}{\sqrt{4\pi}}\right)^{\!\!d}\sum_{n\geq 0} s^{\frac{2n-d}{z}}\frac{\Gamma(\frac{d-2n}{z})}{\frac{z}2\Gamma(\frac{d}{2}-n)}a^{(d)}_{n}\,.
\end{align}
Adding conformal coupling modifies $a^{(d)}_{n}$ but the pre-factor stays  $2\pi \delta(m)\frac{1}{4\pi s} \left(\frac{1}{\sqrt{4\pi}}\right)^{\!\!d}$. Hence, we have the generalised result
\begin{align}
\label{eq:mainresultagain}
\mathcal{A}^{g}_{d+1}= 2\pi \delta(m) \mathcal{A}^{r}_{d+2}
\end{align}
where $\mathcal{A}^{g}_{d+1}$ is the Weyl anomaly of a theory of a single complex scalar field of
charge $m$ under a 
$U(1)$ symmetry  living
in $d+1$ dimensions with dynamical exponent $z$ and $\mathcal{A}^{r}_{d+2}$ is the Weyl anomaly of a field theory living in $d+2$ dimension such that it admits a symmetry under
$t\to \lambda^{z/2}t, x^{d+2}\to \lambda^{z/2}x^{d+2} \text{and\
}x^{i}\to\lambda x^{i}$
for $i=1,\ldots,d+1$.  Thus we have shown that theories with one time
derivative on a time independent curved background do not have any
Weyl anomalies. This is consistent with the perturbative result
obtained previously.

It deserves mention that the operator $\mathcal{M}_{rc;d+2}$ of Eq.~\eqref{curvedoperator} does not
transform homogeneously under Weyl transformations.  In order to construct a Weyl covariant operator
consider generalizing the metric \eqref{eq:metricslice} to the following form
\begin{equation}
ds^2= N dx^+dx^-+h_{ij}dx^idx^j\,.
\end{equation}
If $N$ is independent of $x^-$ the metric for the reduced theory will include a general lapse
function $N$.  Then we replace $(\nabla^2)^{\frac{z}{2}}$ by  $\mathcal{O}^{(d+2z-4)}\mathcal{O}^{(d+2z-8)}\cdots\mathcal{O}^{(d+4)}\mathcal{O}^{(d)}$ with $\mathcal{O}^{(p)}$  defined as
\begin{align}
\mathcal{O}^{(p)}\equiv \nabla^{2}-\frac{p}{4(d-1)}R+ \frac{2+p-d}{z}\frac{\partial_{i}N}{N}h^{ij}\partial_{j}+\frac{d}{4z^{2}}(2+p-d)\frac{\partial_{i}N}{N}h^{ij}\frac{\partial_{j}N}{N}
\end{align}
Under $h_{ij}\to e^{2\sigma}h_{ij}$, $N\to e^{z\sigma}N$ and $\psi \to
e^{-\frac{p}{2}\sigma}\psi$, \bg{this operator transforms covariantly}, in the sense that 
\begin{align}
\mathcal{O}^{(p)}\psi \to e^{-(\frac{p}{2}+2)\sigma}\mathcal{O}^{(p)}\psi\,.
\end{align}
Therefore, under the Weyl rescaling $h_{ij}\to e^{2\sigma}h_{ij}$, $N\to e^{z\sigma}N$ and $\phi \to
e^{-\frac{d}{2}\sigma}\phi$ we have that
\begin{align}
\label{eq:confCoupling}
N\sqrt{h}\,\phi^{*}\mathcal{O}^{(d+2z-4)}\mathcal{O}^{(d+2z-8)}\cdots\mathcal{O}^{(d+4)}\mathcal{O}^{(d)}\phi
\end{align}
is invariant under under Weyl transformations. 

Adding the conformal coupling will modify the expressions for
$a^{(d)}_{n}$, but scaling with respect to $s$ will remain unmodified. Hence
we can enquire about existence or absence of potential Weyl
anomalies. To have a non-vanishing Weyl anomaly, we need to have an
$s$ independent term in the heat kernel expansion. This is possible
only when $\frac{2n-d}{z}=1$, {\it i.e.}, when $d+z$ is even; see Eqs.~\eqref{eq:KMrcexp} and~\eqref{eq:KMgcexp}. Since for
a local Lagrangian $z$ must be even, this condition corresponds to
even~$d$ \footnote{\bg{Giving up on \bg{the} requirement of locality
    allows $z$ to be any positive real number. In \bg{this case}, the anomaly is expected to \bg{be}
    present whenever $d+z$ is even. It might be of potential interest
    to look at these cases carefully and make sure that non-locality
    does not provide any obstruction in the anomaly calculation and
    \bg{that the}  renormalization process can be done in a consistent manner.}}.  This is expected because of the following reason: the
scalars we can construct out of geometrical data (that can potentially appear
as a trace anomaly) have even dimensions and the volume element scales
like $\lambda^{d+z}$, so that in order to form a scale invariant quantity $d+z$
has to be even. Now when $d$ is even, we have $s$
independence for $n=(d+z)/2$ and the coefficient of $s^{0}$ is given by
$\left(\frac{1}{\sqrt{4\pi}}\right)^{d}
(-1)^{1-\frac{z}2}a^{d}_{\frac{d+z}{2}}$. Hence, the result relating anomalies in the parent and
reduced theory, Eq.~\eqref{eq:mainresultagain}, still holds.

\section{Summary, Discussion and Future directions}
\label{conclusion}
We have shown that for a $d+1$ dimensional Schr\"odinger invariant field theory of a
single complex scalar field carrying charge $m$ under $U(1)$ symmetry, the Weyl anomaly,
$\mathcal{A}^{G}_{d+1}$, is given in terms of that of a relativistic free scalar field
living in $d+2$ dimensions, $\mathcal{A}^{R}_{d+2}$, via
\begin{align}
\mathcal{A}^{G}_{d+1}=  2\pi \delta (m) \mathcal{A}^{R}_{d+2}\,.
\end{align}
Here the parent $d+2$ theory lives in a space-time with null isometry generated by the
Killing vector~$\partial_{-}$ so that the metric can be given in terms of a $d+1$
dimensional Newton-Cartan structure. The result is shown to be generalised to
\begin{align}
\mathcal{A}^{g}_{d+1}= 2\pi \delta(m) \mathcal{A}^{r}_{d+2}\,,
\end{align}
where $\mathcal{A}^{g}_{d+1}$ is the Weyl anomaly of a theory of a single complex scalar field of
charge $m$ under an 
$U(1)$ symmetry  living
in $d+1$ dimensions with dynamical exponent $z$, while 
$\mathcal{A}^r_{d+2}$ is the Weyl anomaly of an $SO(1,1)\times SO(d)$
invariant theory living in $d+2$ dimension such that it admits
symmetry under
$t\to \lambda^{z/2}t$, $x^{d+2}\to \lambda^{z/2}x^{d+2}$ and
$x^{i}\to\lambda x^{i}$ for $i=1,\ldots,d+1$.

To obtain  information regarding the anomaly, we introduced a method  to
systematically handle the heat kernel for a theory with kinetic term
involving one time derivative only. We provided crosschecks and
consistency checks on our  heat kernel prescription. 
One may worry  that  to properly define a heat kernel the square of the derivative operator must be considered. This would also be the case for, say, the Dirac operator. In fact, one can properly define it  this way;
see, for example,  Ref.~\cite{Witten:2015aba}. 

The result obtained regarding the anomaly of
  Schr\"odinger field theory is consistent with the one by Jensen
  \citep{Jensen:2014aia}. Auzzi {\it et al},
\citep{Auzzi:2016lxb} have studied the anomaly for a Euclidean operator  given by
\begin{align}
 \mathcal{M}^{\prime}_{E,g}= 2m\sqrt{-\partial_{t}^{2}} - \nabla^{2}\,,
\end{align}
with eigenspectra given by $|\vec{k}|^{2} + 2m |\omega| \geq 0$. 
One can
define the heat kernel for this operator as well, but 
the eigenspectra of this operator is not analytically related to that
of $\mathcal{M}_{M,g}=2\imath m\partial_{t}+\nabla^{2}$, which is
$-k^{2} + 2m \omega$. As a result the propagator in
  $\omega$-$\vec k$ space has a cut on the complex $\omega$ plane with
  branch point at the origin, making the analytic continuation to
  Minkowski space problematic. It is known that the
two point correlator of Schr\"odinger field theory is constrained and
has a particular form as elucidated in
Ref.~\citep{Nishida:2007pj,Goldberger:2014hca}. While our prescription and the
resulting Euclidean correlator conforms to that form, it is not clear
how the Euclidean Schr\"odinger operator defined
in Ref.~\citep{Auzzi:2016lxb} does, if at all.  
Finally, we note that the operator $\sqrt{-\partial_t^2}$ is non-local (in the sense that the
kernel, defined by $\sqrt{-\partial_t^2}f(t)=\int\! dt'\,K(t-t')f(t')$, has non-local support,
$K(t)=2\partial_t\text{P}\frac{1}{t}$). 

There are several avenues of investigation suggested  by this work:
\begin{enumerate}
\item What happens in the case of several scalar fields with different
  charge interacting with each other while preserving Schr\"odinger
  invariance in flat space-time? How is the pre-factor $\delta(m)$
  modified?
\item  It is not obvious
  how null reduction of a theory of a Dirac spinor in $d+2$ dimensions can result in a  Lagrangian in $d+1$
  dimensions of the form
  $\mathcal{L}=2\imath m \psi^{\dagger}\partial_{t}\psi +\psi^{\dagger}\nabla^{2}\psi$, let alone
  one with
  $\mathcal{L}=2\imath m \psi^{\dagger}\partial_{t}\psi -\psi^{\dagger}(-\nabla^{2})^{z/2}\psi$ for
  $z\ne2$. On the other hand, as we have seen, the functional integral over non-relativistic anti-commuting fields
  yields the same determinant as that of commuting fields (only a positive power). Hence,  the
  anomaly of the anti-commuting field is the negative of that of the commuting field. 
\item  Calculations using
  the same Euclidean operator as in Ref.~\citep{Auzzi:2016lxb} give a non-vanishing entanglement
  entropy in the ground state~\citep{Solodukhin:2009sk}. By contrast, for the operator
  $\mathcal{M}_{M,g}=2\imath m\partial_{t}+\nabla^{2}$, the entanglement entropy in the ground state
  vanishes, since for this local non-relativistic field theory $\phi(x)|0\rangle=0$ and hence the
  ground state is a product state. It would be of interest to verify this result by direct
  computation using a method based on  our prescription.
\item The method described in Sec.~\ref{sec:gener} to compute Weyl anomalies in theories with
  $z\ne2$ is not sufficiently general in that, by assuming the metric is time independent and has
  constant lapse, it neglects anomalies involving extrinsic curvature or gradients of the lapse
  function. A future challenge is to develop a more general computational method.

\end{enumerate}

We hope to come back to  these questions in the future.

\begin{acknowledgments}
 SP would like to thank
Mainak Pal, Shauna Kravec and Diptarka Das for useful discussions. \bg{The authors would also like to acknowledge constructive and useful comments \bg{by} the referee.} This work was supported in part by the US
Department of Energy under contract DE-SC0009919.
\end{acknowledgments}

\appendix
\section{Technical Aspects of Heat Kernel for one time derivative theory}\label{ap1}
Here's one more perspective of why $\delta(m)$ appears in heat kernel
for one-time derivative theory using the eigenspectra of the operator
$\mathcal{M}_{g}$ with one time derivative. The Minkowski $\mathcal{M}_{M,g}$ operator is given
by
\begin{align}
\mathcal{M}_{M,g}= 2\imath m \partial_{t} - (-\nabla^2)^{z/2}
\end{align}
and eigenspectra is given by $2m\omega -k^{z}$. Now, we can not
directly define the heat kernel since the eigenvalues range from
$-\infty$ to $\infty$, and therefore it blows up. A similar situation
also arises in relativistic theory where the eigenspectra is given by
$-\omega^{2}+k^{2}$. There we define the heat kernel by Euclideanizing
the time co-ordinate so that the eigenvalues become
$\omega^{2}+k^{2}\geq 0$ and this positive definiteness allows for
convergence. Technically, we can always define heat kernel for an
operator $M$ as long as the eigenvalues of $M$ have positive real
part. Building up on our experience to deal with the relativistic case, we
use analytic continuation here as well. We define the Euclidean
operator as
\begin{align}
\mathcal{M}_{E,g}=2 m \partial_{\tau} + (-\nabla^2)^{z/2}
\end{align}
with eigenspectra given by
$\lambda_{k,\omega}=-2\imath m\omega+k^{z}$. Evidently,
$\text{Re}\left(\lambda_{k,\omega}\right)\geq 0$, hence we have a well  defined  heat kernel,  given by
\begin{align}
K_{\mathcal{M}_{E,g}}=\Tr e^{-s\mathcal{M}_{E,g}} 
= \int \frac{d^{d}k}{(2\pi)^{d}} e^{-sk^{z}} \int \frac{d\omega}{2\pi}  e^{-2m\imath s \omega} =  \frac{\delta (m)}{2 s} \frac{2}{\Gamma(\frac{d}{2})}  \frac{\Gamma \left(\frac{d}{z}+1\right)}{d \left(\sqrt{4\pi s^{\frac{2}{z}}}\right)^{d}}
\end{align}
 
Similarly, the Euclidean heat kernel is well defined for the operator
$\mathcal{M}_{rc;d+2}=
\nabla^{2}_{t,x}-(-\nabla^{2}_{x^{i}})^{z/2}$, where $i=1,2, \ldots d$ and $x\equiv x^{d+2}$.
If we Wick rotate to Euclidean time $\tau$, the eigenvalues
of the operator $\mathcal{M}_{rc;d+2}$ are given by
$\omega^{2}+(k^{d+2})^{2}+(|\vec{k}|^2)^{z/2}\geq 0$. \bg{The
  presence of $\delta(m)$ can more formally be treated with an extra
  regulizer $\eta$, as discussed in the last few paragraphs of
  \ref{sec:heatKdircalcsub} for $z=2$; a similar argument, using the regulator $\eta$, applies to any $z$}.

\def\bdr{\begin{aligned}}
\def\edr{\end{aligned}}

\bg{\section{Riemann normal co-ordinate and coincident limit}
\label{riemann}
In this appendix we show $x^-$ independence of  quantities relevant to the computation of  the
coincidence limit of the Heat Kernel when the light cone reduction technique is used.  
We assume that the daughter theory is coupled to a Newton Cartan
structure, satisfying \bg{the} Frobenius condition, {\it i.e.}, $\vec{n} \wedge d\vec{n}=0$ is satisfied. This condition allows a foliation of the manifold globally. Thus, without loss of generality, the metric is given by 
\begin{align}
\bdr
g_{\mu\nu}&=n_{\mu}n_{\nu}+h_{\mu\nu}\\
n_{\mu}&=(n,0,0,\cdots, 0)\ , \quad h_{\tau \nu}=0.
\edr
\end{align}
Using \eqref{eq:Milneconds} and the fact $h_{ij}$ is a positive definite matrix, we thus have
\begin{align}
h^{\tau\nu}=0\ , \quad v^{\mu}=\left(\tfrac{1}{n},0,0,\cdots, 0\right).
\end{align}
The form of the metric, \bg{to} which the reduced theory is coupled, corresponds to a
parent space-time metric $G_{MN}$, \bg{with non-vanishing components} given by
\begin{align}\label{schoice}
\bg{ G_{-+}=n\,, \quad G_{ij}=h_{ij}\,.}
\end{align}
\bg{In addition, we assume that} the parent space-time admits a
null isometry \bg{so that  $h_{ij}$ and $n$ are independent of $x^{-}$. }

In what follows, we will work with this particular choice of metric $G_{MN}$ \bg{\eqref{schoice}}. Without loss of generality, we choose $x_1=(0,0,\cdots,0)$ (we call it point $P$) and construct the Riemann normal co-ordinate with the origin as the base point. The Riemann normal co-ordinate $y^{M}$, is given in terms of the original co-ordinate $x^{M}$ as follows~\citep{Brewin:2009se}:
\begin{align}\label{expansion}
y^{M}= x^{M}+ f^{M}_{AB}x^{A}x^{B}+f^{M}_{ABC}x^{A}x^{B}x^{C}+\cdots\,,
\end{align}
where the index $M$ runs over $+,-,1,2,3,\cdots,d$. In \bg{the}
coincident limit of \bg{the}  reduced theory, {\it i.e.}, $x_{2}^{\mu}
\to 0 $, for $\mu =+,1,2,\cdots, d$ (\bg{with} $x_{2}^{-}$ \bg{possibly} different from $0$), we claim that
\begin{align}\label{co}
\left[y_2^{\mu}\right]=0, \qquad \left[y_2^{-}\right]=x_2^{-},
\end{align}
where  \bg{henceforth}  the square bracket is used to denote the coincident limit in \bg{the} reduced theory.

We note that \bg{$[f^{M}_{ABC...}x^{A}x^{B}x^{C}\cdots]=0$} whenever any of the indices is not
$-$.  \bg{R}ecall that $f^{M}_{ABC\cdots}$ are
constructed out of derivatives acting on metric. Thus, $f^{M}_{\underbrace{--\cdots-}_{N\
    \text{indices}}}$ can be non-zero only if \bg{it contains} $N$ 
\bg{factors} of \bg{the} metric tensor $G_{-K_{i}}$, where $K_{i}$ is a running index with
$i=1,2,\cdots, N$. This is because $G_{--}=0$ and derivatives can not carry \bg{the} ``$-$" index as
\bg{the} metric components are $x^{-}$-independent. Moreover, \bg{by dimensional analysis}
$f^{M}_{\underbrace{--\cdots-}_{N}}$ has $N-1$ derivatives\bg{
  $f^{M}_{\underbrace{--\cdots-}_{N}}$}. Schematically, this assumes one of the following forms
\begin{align}
\partial_{A_1}\cdots \partial_{A_{N-1}} G_{-K_1}\cdots G_{-K_N} G^{MA_i} G^{A_{i_1} K_{j_1}}G^{A_{i_2}A_{j_2}}\cdots G^{K_{i_3}K_{j_3}}\bg{\cdots}\,,\\
\partial_{A_1}\cdots \partial_{A_{N-1}} G_{-K_1}\cdots G_{-K_N} G^{MK_i} G^{A_{i_1} K_{j_1}}G^{A_{i_2}A_{j_2}}\cdots G^{K_{i_3}K_{j_3}}\bg{\cdots}\,.
\end{align}
Here the derivatives are assumed to act on all possible combinations, resulting in different
possible terms. For example, for $N=2$, one can have \bg{the following} terms:
\begin{equation}
\label{caseN2}
\begin{aligned}
G^{MA_1}G^{K_1 K_2}G_{-K_2}\partial_{A_1}G_{-K_1}\,,\\
G^{MK_2}G^{A_1 K_1}G_{-K_1}\partial_{A_1}G_{-K_2}\,,\\
G^{MK_2}G^{A_1 K_1}G_{-K_2}\partial_{A_1}G_{-K_1}\,.
\end{aligned}
\end{equation}
\bg{There can not be any $x^-$ derivative for a term to be non-vanishing.} \bg{This implies the indices $A_i$
  are contracted among themselves, except possibly for one contracted with $G^{MA_i}$, and the
  indices $K_i$ are contracted among themselves. But since $G_{-K}=0$ except for $G_{-+}$, and $G^{++}=0$, any term for which two factors of the metric tensor,  $G_{-K_{i_1}}$ and
$G_{-K_{i_2}}$, are contracted via $G^{K_{i_1}K_{i_2}}$ vanish. 
}

Next, we show that $\left[\Delta_{VM}\right]=1$. The expression for $\Delta_{VM}$\bg{, Eq,~\eqref{eq:vandet},}
involves bi-derivatives of \bg{the geodetic interval, Eq.~\eqref{eq:geodeticinterval}, and the }
determinant of \bg{the} metric. To begin with, we turn our attention to \bg{the} determinant of \bg{the} metric and note that
\begin{align}
\left[G^{\prime}(y_2)\right]=J^2(0,x_2^{-},0,\cdots,0)G(0,x_2^{-},0,\cdots,0)\,,
\end{align}
where \bg{a} prime indicates quantities in Riemann normal co-ordinate and $J$ is the Jacobian associated with \bg{the} co-ordinate transformation~\eqref{expansion}. The $x^{-}$ independence in the original co-ordinate guarantees that $G(0,x_2^{-},0,\cdots,0)=G(0,0,0,\cdots,0)$, hence we have 
\begin{align}\label{a1}
\left[G^{\prime}(y_2)\right]=\left(\frac{J(0,x_2^{-},0,\cdots,0)}{J(0,0,0,\cdots,0)}\right)^{2}G^{\prime}(0).
\end{align}
\bg{Next consider the geodetic interval}  from point
$P$ to point $Q$\bg{.}
\bg{In} Riemann normal co-ordinates~\citep{Brewin:2009se}
\begin{align}
\bg{y^{M}_{2}=y^M(Q)}=y^{M}_{1}+s_{Q}\frac{dx^{M}}{ds}\bigg|_{s=0}\,,
\end{align}
where $s_{Q}$ is the value of \bg{the} affine parameter at  $Q$ and \bg{ $s=0$} at $P$, \bg{with
  $y^{M}_{1}=y^M(P)$}. \bg{Using Eq.~\eqref{eq:geodeticinterval}, hence we have}
\begin{align}
2\sigma(y_2,y_1) = G_{MN}(0)(y_2^{M}-y_{1}^{M})(y_2^{N}-y_{1}^{N})= G^{\prime}_{MN}(0)(y_2^{M}-y_{1}^{M})(y_2^{N}-y_{1}^{N})
\end{align}
\bg{where we have used $G^{\prime}_{MN}(0)=G_{MN}(0)$.} \bg{It follows that}
\begin{align}
\label{a2}
\Delta_{VM}= \left(\frac{G^{\prime}(y_2)}{G^{\prime}(0)}\right)^{-1/2}\,.
\end{align}
\bg{We have} continued back to Minkowskian signature (the definition in
\bg{Eq.}~\eqref{eq:vandet} is for metric with Euclidean signature). \bg{Since} $\Delta_{VM}$ is a bi-scalar, use \bg{of Eqs.~}\eqref{a1}
\bg{and}~\eqref{a2} \bg{and of  $J(0,0,0,\cdots,0)=1$} \bg{gives}
 \begin{align}\label{coin}
 \left[\Delta_{VM}\right]=\left(\frac{J(0,x_2^{-},0,\cdots,0)}{J(0,0,0,\cdots,0)}\right)^{-1}=J^{-1}(0,x_2^{-},0,\cdots,0)
\end{align}
 in the original co-ordinate \bg{system,} $x^{M}$.
 \bg{Equation}~\eqref{coin} is consistent with the result that $\Delta_{VM}=1$ when all the
 co-ordinates, including $x^{-}$, coincide, {\it i.e.}, when $x_{2}^{-}=0$.

We aim to show that 
\begin{align}
\left[\det\bigg(\frac{\partial y^{M}}{\partial x^{N}}\bigg)\right]= \det\bigg(\left[\frac{\partial y^{M}}{\partial x^{N}}\right]\bigg) = 1
\end{align}
From Eq.~\eqref{expansion} we have
\begin{align}
\
\left[\frac{\partial y^{M}}{\partial x^{N}}\right] = \delta^{M}_{N}+(f^{M}_{N-}+f^{M}_{-N})x^{-}+(f^{M}_{N--}+f^{M}_{-N-}+f^{M}_{--N})x^{-}x^{-}+\cdots
\end{align}
Consider first the lowest two terms in the expansion. Explicitly, we have~\citep{Brewin:2009se}
\begin{equation}
2f^{M}_{N-}=2f^{M}_{-N}=\Gamma^{M}_{N-}=-\frac{1}{2}G^{Mi}\partial_{i}G_{N-}-\frac{1}{2}G^{M+}\partial_{+}G_{N-}+\frac{1}{2}G^{M+}\partial_{N}G_{+-}\,.
\end{equation}
It follows that $f^{M}_{N-} \neq 0$ only for \bg{$M=-$ or $N=+$}. Similarly, \bg{$f^{M}_{(N--)}\ne0$
  provided $M=-$ or $N=+$ }, since~\citep{Brewin:2009se}
\begin{align}
6f^{M}_{NIJ}=\Gamma^{M}_{NE}\Gamma^{E}_{IJ}+\partial_{N}\Gamma^{M}_{IJ}
\end{align}
\bg{By an argument analogous to that below Eqs.~\eqref{caseN2} one can show that
$[f^{M}_{N--\cdots-}]=0$ (at least three $-$ subscripts).}
%
\bg{Schematically}
\[
\left[\bigg(\frac{\partial y^{M}}{\partial x^{N}}\bigg)\right]= 
\begin{pmatrix}
1 & * & * & \dots & \dots &*\\
0 & 1 & 0 & \dots & \dots & 0\\
0 & * & 1 & 0 & \dots & 0\\
\vdots & \vdots &  &\ddots & \ddots &\\
0 & * & 0 & 0& 1 & 0\\
0 & * & 0 & 0& 0 &1
\end{pmatrix}
\]
\bg{where a ``$*$'' means a non-zero entry.}
Thus, the \bg{matrix has unit} determinant and we have, using Eq.~\eqref{coin},
\begin{align}
\label{eq:deltais1}
\left[\Delta_{VM}\right]=1\,.
\end{align}

\bg{Lastly, we turn to the heat
kernel expansion coefficients,  $a_n$. They are determined by the recursive relation ~\cite{Mukhanov:2007zz},
\begin{align}
n a_n+\partial_{M}\sigma\partial^{M}a_n=-\Delta^{-1/2}_{VM}\mathcal{M}\left(\Delta^{1/2}_{VM} a_{n-1}\right)\,,
\end{align}
and $a_0=1$, where $\mathcal{M}$ is the relativistic operator in the parent theory. The condition of
$x^-$ independence of $[a_n]$, $[\partial_i a_n]$ and $[\partial_i\partial_j a_n]$ can be imposed on
the recursion self-consistently. To show this one uses  $x^-$
independence of  $\left[\Delta_{VM}\right]$, $\left[\partial_i\Delta_{VM}\right]$ and
$\left[\partial_i\partial_j\Delta_{VM}\right]$, which follows from an argument
similar to the one used to establish Eq.~\eqref{eq:deltais1} 
}
}

\bg{\section{\bg{Explicit Perturbative Calculation of $\eta$-regularized Heat Kernel}}
\label{hketa}
In this appendix \bg{we give an explicit perturbative computation that shows the vanishing of
  the anomaly for  a class of curved backgrounds. This serves to verify the general arguments presented in the body of the manuscript in  a specific, simple example, and allows us to study explicitly  the $\eta$ regulated Heat Kernel asking in particular whether  the 
  $\eta\to0$ limit is a well defined limit as  $m\neq 0$.} To be specific, we
  compute the heat kernel on a curved background, characterized by
\begin{align}
n_{\mu}=\left(\frac{1}{1-n(x)},0,0\right), &\qquad v^{\mu}=\left(1-n(x),0,0\right)\\
 \quad h_{ij}=\delta_{ij}\, , &\quad \sqrt{g}=\sqrt{\text{det}(n_{\nu}n_{\nu}+h_{\mu}h_{\nu})}=\frac{1}{1-n(x)}.
\end{align}
where $n(x)$ is a function of space only and $h_{i0}=0$. The special choice is inspired by
\cite{Auzzi:2016lxb}and additionally serves the purpose of affording a direct comparison
with that work. We will perform a perturbative calculation as an expansion in $n(x)$.  We
will specialize to a $2+1$ dimensional Schr\"odinger field theory coupled to this
background. The action is given by
\begin{align}\label{Action}
S=\int dtd^2x\ \bg{N}\left(2m\phi^{\dagger}\imath\tfrac{1}{N}\partial_t\phi-h^{ij}\partial_i\phi^{\dagger}\partial_j\phi-\xi R\phi^{\dagger}\phi\right)
\end{align}
where $N(x)=\tfrac{1}{1-n(x)}$ and $R$ is the Ricci scalar of the $3+1$ dimensional geometry, on which the parent theory lives. 

As we will see, the result of this calculation is that the Weyl anomaly,
corresponding to the theory described by Eq.~\eqref{Action} is given by
\begin{align}
\label{eq:app:A_G}
\mathcal{A}_G=2\pi\delta(m) \left(-aE_4+cW^2+bR^2+dD_MD^MR\right)
\end{align}
where the coefficients $a,b,c,d$ are given by: 
\begin{equation}
\label{ANOMALY}
\begin{aligned}
a&=\frac{1}{8\pi^2}\frac{1}{360}\ , \quad b=\frac{1}{8\pi^2}\frac{1}{2}\left(\xi-\frac{1}{6}\right)^2\ ,\\
c&=\frac{1}{8\pi^2}\frac{1}{120}\ , \quad d=\frac{1}{8\pi^2}\left(\frac{1-5\xi}{30}\right)\ .
\end{aligned}
\end{equation}

These are exactly the same as in the expression for the Weyl Anomaly of a relativistic complex scalar field
theory\footnote{\bg{The Weyl anomaly of a complex scalar field is twice of that of a real scalar field.}} living in one higher dimension~\citep{Deser:1976yx,Brown:1976wc,Dowker:1976zf,Hawking:1976ja,Christensen:1977jc,Duff:1977ay,Duff:1993wm}:
\begin{align}
\mathcal{A}_R= \left(-aE_4+cW^2+bR^2+dD_MD^MR\right)\,.
\end{align}

To arrive at this result, we proceed by considering the heat kernel of the following Euclidean operator, corresponding to the action in Eq.~\eqref{Action}, namely
\begin{align}
\mathcal{M}_{E,c}=2m\tfrac{1}{N}\partial_\tau - \mathcal{D}^2+\xi R\,,
\end{align}
where we have 
\begin{align}
\mathcal{D}^2 &= \frac{1}{\sqrt{g}}\partial_{i}\left(\sqrt{g}h^{ij}\partial_j\right)= \partial^2 + \left(1+n\right)\left(\partial_i n\right)\partial_i\,,\\
R &= -2\partial^{2}n-2n\partial^2n-\frac{7}{2}\partial_i n \partial_i n+\cdots\,,\\
-g^{1/4}\mathcal{D}^2\left(g^{-1/4}\delta(x)\right)&= -\partial^2 \delta(x) +\delta(x) \left(\frac{1}{2}\partial^2n+\frac{1}{2}n\partial^2n+\frac{3}{4}\partial_in\partial_in\right).
\end{align}

The Euclidean operator can be expressed as the one in flat space-time, perturbed by the background field $n(x)$:
\begin{align}\label{pert}
\nonumber \langle \vec{x},\tau|\mathcal{M}_{E,c}|\vec{x}^{\prime},\tau^{\prime}\rangle&=\langle \vec{x},\tau|\mathcal{M}_{E,f}|\vec{x}^{\prime},\tau^{\prime}\rangle+mP_{1}(x)\partial_\tau\delta(\vec{x}-\vec{x}^{\prime})\delta(\tau-\tau^{\prime})\\
& \quad+P_2(x)\delta(\vec{x}-\vec{x}^{\prime})\delta(\tau-\tau^{\prime})\,,
\end{align}
where the subscript $c$ and $f$ denote the curved and flat space-time respectively while $E$ denote the Euclidean nature of the operator. Here we have introduced
\begin{align}\label{P}
P_1(x)=2n(x), \quad P_2(x)= \left(\frac{1}{2}\partial^2n+\frac{1}{2}n\partial^2n+\frac{3}{4}\partial_i n\partial_i n\right)-\xi \left(2\partial^2 n+2n\partial^2 n+\frac{7}{2}\partial_i n\partial_i n\right).
\end{align}

The heat kernel can be obtained as a perturbative expansion of the background fields as follows:
\begin{align}
K(s)=\exp\left[-s\left(\mathcal{M}_{E,f}+P\right)\right]=\sum_{N=0}^{\infty}(-1)^{N}K_{N}(s)\,.
\end{align}
The $K_{N}(s)$ is defined as follows:
\begin{align}\label{Kn}
K_{N}(s)=\int_0^s\!\!ds_{N} \int_0^{s_{N}}\!\!\! ds_{N-1}\cdots \int_0^{s_{2}}\!\!\! ds_{1}\ G(s-s_N)PG(s_N-s_{N-1})P\cdots G(s_2-s_1)P G(s_1)\,.
\end{align}
where $G(s)=e^{-s\mathcal{M}_{E,f}}$ and $P$ is the perturbation~\eqref{pert}, explicitly given by
\begin{align}
\langle \vec{x},\tau|P|\vec{x}^{\prime},\tau^{\prime}\rangle&=mP_{1}(x)\partial_\tau\delta(\vec{x}-\vec{x}^{\prime})\delta(\tau-\tau^{\prime})+P_2(x)\delta(\vec{x}-\vec{x}^{\prime})\delta(\tau-\tau^{\prime}).
\end{align}
One can now complete the calculation by using the matrix element of $G(s)$ as given by 
\begin{align}\label{fhk}
\nonumber \mathcal{G}_{g,E}\left(s; (\vec{x}_2, \tau_2),(\vec{x}_1, \tau_1)\right)&\equiv \langle \vec{x}_2, \tau_2|G(s)|\vec{x}_1,\tau_1\rangle \\
& = \frac{1}{\pi}\left(\frac{1}{4\pi s}\right)^{d/2}\left[\frac{s\eta}{\left(2ms-\tau_2+\tau_1\right)^2+s^2\eta^2}\right]e^{-\frac{(\vec{x}_2-\vec{x}_1)^2}{4s}},
\end{align}
which corresponds to the heat kernel expression for the $\eta$-regulated Euclidean operator:
$\mathcal{M}^{\prime}_{E,g}=2m\partial_{\tau}-\nabla^2+\eta\sqrt{-\partial^{2}_{\tau}}$, as
discussed in the last few paragraphs of \ref{sec:heatKdircalcsub}.\footnote{\bg{In curved space-time,
  $\mathcal{M}^\prime_{E,g}$ includes a perturbation $n(x)\eta\sqrt{-\partial^{2}_{\tau}}$, that,
  however, does not contribute to the anomaly  in the $\eta\to0$ limit. This term's contribution to
  $K_{1}$ is proportional to $\frac{\eta\left(\eta^2-4m^2\right)}{\left(\eta ^2+4 m^2\right)^2}$
  that vanishes as $\eta\to 0$, without giving a $\delta(m)$ (or any derivative of
  $\delta(m)$). This term's contributions to  $K_2$ also vanish as $\eta\to 0$. We omit these terms for simplicity for rest of the appendix.}} This reproduces Eq.~\eqref{heatkernelg00} as $\eta\to0$. 

The evaluation of Eq.~\eqref{Kn} follows the procedure sketched out in the appendix of
\cite{Auzzi:2016lxb}. We separate the contributions from $P_1$
and $P_2$ to $K_1$ as follows:
\begin{align}
K_{1P_1}(s)&=\left(\frac{\tfrac{\eta}{2}}{m^2+\tfrac{\eta^2}{4}}\right)\left(\frac{-1}{4m^2+\eta^2}\right)\frac{8m^2}{\left(4\pi s\right)^{2}}\left(P_1+\tfrac{s}{6}\partial^2P_1+\tfrac{s^2}{60}\partial^2\partial^2P_1+\cdots\right),\\
K_{1P_2}(s)&=\left(\frac{\tfrac{\eta}{2}}{m^2+\tfrac{\eta^2}{4}}\right) \frac{2}{\left(4\pi s\right)^{2}}\left(sP_2+\tfrac{s^2}{6}\partial^2P_2+\cdots\right), 
\end{align}
and for  $K_2$, which  gets contributions quadratic in $P_1$ and $P_2$, as follows:
\begin{align}
\nonumber K_{2P_1P_1}(s)&=\frac{\left(24 m^2-2 \eta ^2\right)}{\left(\eta ^2+4 m^2\right)^2}\left(\frac{2m^2\eta}{4m^2+\eta^2}\right)\frac{1}{(4\pi s)^2}\bigg(P_1^2+\tfrac{s}{3}P_1\partial^2P_1+\tfrac{s}{6}\partial_iP_1\partial_iP_1\\
&\quad+\frac{s^2}{180}\left(6P_1\partial^2\partial^2P_1+5\partial^2P_1\partial^2P_1+12\partial_i\partial^2P_1\partial_iP_1+4\left(\partial_i\partial_jP_1\right)\left(\partial_i\partial_jP_1\right)\right)\bigg)\\
K_{2P_1P_2}(s)&=\left(\frac{\tfrac{\eta}{2}}{m^2+\tfrac{\eta^2}{4}}\right)\left(\frac{-1}{4m^2+\eta^2}\right)\frac{8m^2}{\left(4\pi s\right)^{2}}\Big(\tfrac{s}{2}P_1P_2\nonumber\\
&\qquad\qquad+\tfrac{s^2}{12}(P_2\partial^2P_1+P_1\partial^2P_2+\partial_iP_1\partial_iP_2)+\cdots\Big)\\
K_{2P_2P_1}(s)&=K_{2P_1P_2}(s)\\
K_{2P_2P_2}(s)&=\left(\frac{\tfrac{\eta}{2}}{m^2+\tfrac{\eta^2}{4}}\right) \frac{2}{\left(4\pi s\right)^{2}}\left(\tfrac{s^2}{2}P_2^2+\cdots\right)
\end{align}

The anomaly is determined by the $s$-independent terms in $K_N$.  In $\eta\to 0$ limit, factors of  $\delta(m)$ arise, after use of  the  following easily verifiable limits
\begin{align*}
\lim_{\eta\to 0} &\ \left(\frac{\tfrac{\eta}{2}}{m^2+\tfrac{\eta^2}{4}}\right)\left(\frac{8m^2}{4m^2+\eta^2}\right)= \pi \delta(m)\ ,\\
\lim_{\eta\to 0}&\ \left(\frac{\tfrac{\eta}{2}}{m^2+\tfrac{\eta^2}{4}}\right)=\pi\delta(m)\ ,\\
\lim_{\eta\to 0}&\ \frac{24 m^2-2 \eta ^2}{\left(\eta ^2+4 m^2\right)^2}\left(\frac{2\eta m^2}{m^2+\tfrac{\eta^2}{4}}\right)=2\pi \delta(m).
\end{align*} 
In $\eta\to 0$ limit, the $s$ independent terms are given by
\begin{align*}
K_{1P_1}(s)&\ni\frac{\delta(m)}{16\pi}\ \left[-\frac{1}{30}\partial^2\partial^2n\right] \ ,\\
\nonumber K_{1P_2}(s)&\ni\frac{\delta(m)}{16\pi}\ \left[\frac{1}{3}\partial^2P_2\right]\\
\nonumber &=\frac{\delta(m)}{16\pi}\ \bigg[\frac{1}{3}\bigg(\left(\frac{1}{2}-2\xi\right)\partial^2\partial^2n+\left(\frac{1}{2}-2\xi\right)\partial^2n\partial^2n+\left(\frac{1}{2}-2\xi\right)n\partial^2\partial^2n\\
&\quad +\left(\frac{5}{2}-11\xi\right)\partial_in\partial_i\partial^2n+\left(\frac{3}{2}-7\xi\right)\left(\partial_i\partial_jn\right)\left(\partial_i\partial_jn\right)\bigg)\bigg] \ ,\\
K_{2P_1P_1}&\ni\frac{\delta(m)}{16\pi}\ \left[\frac{1}{90}\left(6n\partial^2\partial^2n+5\partial^2n\partial^2n+12\partial_i\partial^2n\partial_in+4\left(\partial_i\partial_jn\right)\left(\partial_i\partial_jn\right)\right)\right] \ ,\\
\nonumber K_{2P_1P_2}+K_{2P_1P_2}&\ni\frac{\delta(m)}{16\pi}\ \left[\frac{-1}{3}(P_2\partial^2n+n\partial^2P_2+\partial_in\partial_iP_2)\right]\\
&=\frac{\delta(m)}{16\pi}\ \left[\frac{-1}{3}\left(\frac{1}{2}-2\xi\right)(\partial^2n\partial^2n+n\partial^2\partial^2n+\partial_in\partial_i\partial^2n)\right]\ ,\\
K_{2P_2P_2}&\ni\frac{\delta(m)}{16\pi}\ \left[P_2^2+\cdots\right] =\frac{\delta(m)}{16\pi}\ \left[\left(\frac{1}{2}-2\xi\right)^2\partial^2n\partial^2n+\cdots\right]\ .
\end{align*}

Using
\begin{align}
R &= -2\partial^{2}n-2n\partial^2n-\frac{7}{2}\partial_i n \partial_i n+\cdots\ ,\\
R^2 &= 4(\partial^{2}n)^2+\cdots\ ,\quad W^2 =\frac{1}{3}(\partial^{2}n)^2+\cdots\ ,\\
E_4 &=2(\partial^{2}n)^2-2(\partial_{i}\partial_{j}n)(\partial_i\partial_jn)+\cdots\ ,\\
D_MD^MR&=-2\partial^4n-2(\partial^{2}n)^2-2n\partial^4n-13(\partial_j n)(\partial_j \partial^2n)-7(\partial_{i}\partial_{j}n)(\partial_i\partial_jn)+\cdots\ .
\end{align}
one  verifies the anomaly expression in Eqs.~\eqref{eq:app:A_G} and \eqref{ANOMALY}. 
Since our calculation only fixes the value of $12b+c$, in oder to break the degeneracy we use
the fact that for $\xi=\frac{1}{6}$ the Wess-Zumino
  consistency condition precludes an $R^2$ anomaly~\cite{Auzzi:2016lxb} and assume $c$ is $\xi$-independent.

We emphasize that the calculation carried out here does not rely on any null cone
reduction technique, hence, this lends further credence to the LCR prescription, which has
correctly produced the $\delta(m)$ factor, as elucidated before.  
}

\bibliographystyle{apsrev4-1}
\bibliography{refs}
\end{document}